\documentclass[11pt]{article}
\pdfoutput=1

\usepackage{jheppub}
\usepackage{amssymb,amsfonts,amsmath}
\usepackage{enumerate}
\usepackage{mathrsfs}
\usepackage{tikz}
\usepackage[section]{placeins}

\newcommand{\be}{\begin{equation}}
\newcommand{\ee}{\end{equation}}
\newcommand{\bea}{\begin{eqnarray}}
\newcommand{\eea}{\end{eqnarray}}

\newcommand{\corr}[1]{\langle{#1}\rangle}

\newcommand{\CD}{\mathcal{D}}

\newcommand{\CI}{\mathcal{I}}
\newcommand{\CJ}{\mathcal{J}}

\newcommand{\CO}{\mathcal{O}}

\newcommand{\CT}{\mathcal{T}}

\newcommand{\CV}{\mathcal{V}}

\newcommand{\CZ}{\mathcal{Z}}

\newcommand{\bK}{{\bf K}}
\newcommand{\bk}{{\bf k}}
\newcommand{\bp}{{\bf p}}
\newcommand{\bq}{{\bf q}}

\newcommand{\bx}{{\bf x}}
\newcommand{\by}{{\bf y}}

\newcommand{\bV}{{\bf V}}

\newcommand{\lr}{\left (}
\newcommand{\rr}{\right )}
\newcommand{\ls}{\left [}
\newcommand{\rs}{\right ]}
\newcommand{\lc}{\left \{}
\newcommand{\rc}{\right \}}

\newcommand{\oln}{\overline}

\newcommand\qt\tau

\newcommand{\p}{\partial}

\renewcommand{\oln}[1]{\overline{#1}}
\renewcommand{\tilde}[1]{\widetilde{#1}}

\makeatletter
\renewcommand{\@seccntformat}[1]{\csname the#1\endcsname.\,\,}
\makeatother

\usepackage{tikz}
\usetikzlibrary{arrows,decorations.pathmorphing,backgrounds,positioning,fit,petri,automata,shadows,calendar,mindmap,decorations.markings,calc}

\makeatletter
\pgfdeclaredecoration{gluon}{coil}
{
  \state{coil}[switch if less than=%
    0.5\pgfdecorationsegmentlength+
    \pgfdecorationsegmentaspect\pgfdecorationsegmentamplitude+%
    \pgfdecorationsegmentaspect\pgfdecorationsegmentamplitude to last,
               width=+\pgfdecorationsegmentlength]
  {
    \pgfpathcurveto
    {\pgfpoint@oncoil{0    }{ 0.555}{1}}
    {\pgfpoint@oncoil{0.445}{ 1    }{2}}
    {\pgfpoint@oncoil{1    }{ 1    }{3}}
    \pgfpathcurveto
    {\pgfpoint@oncoil{1.555}{ 1    }{4}}
    {\pgfpoint@oncoil{2    }{ 0.555}{5}}
    {\pgfpoint@oncoil{2    }{ 0    }{6}}
    \pgfpathcurveto
    {\pgfpoint@oncoil{2    }{-0.555}{7}}
    {\pgfpoint@oncoil{1.555}{-1    }{8}}
    {\pgfpoint@oncoil{1    }{-1    }{9}}
    \pgfpathcurveto
    {\pgfpoint@oncoil{0.445}{-1    }{10}}
    {\pgfpoint@oncoil{0    }{-0.555}{11}}
    {\pgfpoint@oncoil{0    }{ 0    }{12}}
  }
  \state{last}[next state=final]
  {
    \pgfpathcurveto
    {\pgfpoint@oncoil{0    }{ 0.555}{1}}
    {\pgfpoint@oncoil{0.445}{ 1    }{2}}
    {\pgfpoint@oncoil{1    }{ 1    }{3}}
    \pgfpathcurveto
    {\pgfpoint@oncoil{1.555}{ 1    }{4}}
    {\pgfpoint@oncoil{2    }{ 0.555}{5}}
    {\pgfpoint@oncoil{2    }{ 0    }{6}}
  }
  \state{final}{}
}

\def\pgfpoint@oncoil#1#2#3{%
  \pgf@x=#1\pgfdecorationsegmentamplitude%
  \pgf@x=\pgfdecorationsegmentaspect\pgf@x%
  \pgf@y=#2\pgfdecorationsegmentamplitude%
  \pgf@xa=0.083333333333\pgfdecorationsegmentlength%
  \advance\pgf@x by#3\pgf@xa%
}
\makeatother

\newcommand{\qn}{\nu}
\newcommand{\qq}{q}
\newcommand{\bqq}{\mathbf{q}}

\newcommand{\qp}{p}
\newcommand{\bqp}{\mathbf{p}}

\newcommand{\qk}{k}

\newcommand{\qku}{k_1}

\newcommand{\qkd}{k_2}

\let \savenumberline \numberline
\def \numberline#1{\savenumberline{#1.}}

\makeatletter
\def\@fpheader{\relax}
\makeatother

\def\bea{\begin{eqnarray}}
\def\eea{\end{eqnarray}}

\title{\ \vspace{1.6cm} \\
Nonrelativistic Yang-Mills Theory for a Naturally Light Higgs Boson}
\author{Laure Berthier${}^{a}$,
Kevin T. Grosvenor${}^a$, and Ziqi Yan${}^{b,c,d}$}
\emailAdd{berthier@nbi.ku.dk}
\emailAdd{kevin.grosvenor@nbi.ku.dk}
\emailAdd{\\ \qquad\quad\, zyan@perimeterinstitute.ca}
\affiliation{${}^a$Niels Bohr Institute, University of Copenhagen\\
Blegdamsvej 17, DK-2100 Copenhagen \O, Denmark\medskip\\
${}^b$Perimeter Institute for Theoretical Physics\\
31 Caroline St N, Waterloo, ON N2L 6B9, Canada\medskip\\ 
${}^c$Berkeley Center for Theoretical Physics and Department of Physics\\
University of California, Berkeley, CA 94720-7300, USA\medskip\\
${}^d$Theoretical Physics Group, Lawrence Berkeley National Laboratory\\
Berkeley, CA 94720-8162, USA}
\abstract{We continue the study of the nonrelativistic short-distance completions of a naturally light Higgs, focusing on the interplay between the gauge symmetries and the polynomial shift symmetries. We investigate the naturalness of nonrelativistic scalar quantum electrodynamics with a dynamical critical exponent $z=3$ by computing leading power law divergences to the scalar propagator in this theory. We find that power law divergences exhibit a more refined structure in theories that lack boost symmetries. Finally, in this toy model, we show that it is possible to preserve a fairly large hierarchy between the scalar 
mass and the high energy naturalness scale across 7 orders of magnitude, while accommodating a gauge coupling of order 0.1. 
}

\begin{document}

\maketitle

\section{Introduction}
The discovery of the Higgs boson at the LHC \cite{atlas,cms} and the subsequent lack of new resonances suggest that the Standard Model may be self-contained up to a very high energy scale. This possibility puts a new emphasis on the Higgs mass hierarchy problem, which constitutes one of today's most intriguing puzzles of naturalness along with the cosmological constant problem.  

In the past few years, we have seen some interesting surprises with naturalness in the context of nonrelativistic theories \cite{hahi,cmu,snn, nrn, liberatiyukawa}. Recently in \cite{nrn}, a new method was proposed to open up a mass hierarchy for a fundamental scalar by considering a high-energy crossover to nonrelativistic physics, where the Higgs boson exhibits higher-order dispersion relations. In the simplest ``$10$-$20$-$30$" scenario of the mechanism proposed in \cite{nrn},
a hierarchy of $15$ orders of magnitude between the Higgs mass $m$ and the naturalness scale $M$ was achieved. The model also accommodates
the Higgs nonderivative quartic self-coupling $\lambda_h \sim 1$ and 
the Yukawa couplings in the range of
$y_f \lesssim 1$. Despite these successes, after gauging, a simple analysis showed that the ``$10$-$20$-$30$" model predicts unrealistically small gauge couplings and hence small $W$ and $Z$ boson masses. 
However, as noted in \cite{nrn}, this preliminary conclusion about the gauge couplings in this model comes from the most conservative estimates of the quantum corrections, which ensures technical naturalness but does not necessarily optimize it. Moreover, the proposed short-distance completion of a naturally light Higgs involves higher derivative terms. Covariantly coupling such a nonrelativistic scalar field theory to gauge bosons naturally leads to a plethora of interaction terms, which could in principle provide enough room to improve the naive naturalness bounds. In any case, it is clear that a systematic investigation of technical naturalness in nonrelativistic systems with gauge symmetries is needed.

In this paper, we continue the study initiated in \cite{nrn}, with a focus on gauge symmetries. Instead of working with the non-Abelian Yang-Mills gauge group product $SU(3)_C \times SU(2)_L \times U(1)_Y$ of the Standard Model with the usual Higgs doublet, we will focus on nonrelativistic scalar quantum electrodynamics (QED) with a fundamental complex scalar \cite{Gomesgauge, Farakosgauge}. This toy model already allows us to estimate the sizes of various quantum corrections in the Standard Model. 

We will require the systems to possess the ``Aristotelian spacetime symmetries," first discussed in \cite{Penrose68} and then reintroduced in \cite{nrr, nrn}. The Aristotelian spacetime is defined as $\mathbb{R}^{3+1}$ with the flat metric and the preferred foliation by constant time slices. The Aristotelian symmetries contain spatial rotations and translations and time translation, but no boosts (neither Lorentzian nor Galilean). Such spacetimes emerge naturally in the context of nonrelativistic gravity \cite{mqc, qglp}, as the ground-state solutions of the theory with zero cosmological constant.

This paper uses the guiding principle of naturalness. The two main naturalness criteria are \textit{technical naturalness}, as formulated by 't Hooft~\cite{th}, and a stronger concept of naturalness due to Dirac~\cite{DiracN0,DiracN}. The Dirac naturalness criterion states that there should be no unexplained small parameters in a fundamental theory. In this paper, we do not necessarily explain naturalness in the Dirac sense. Instead, we will take 't Hooft's perspective on technical naturalness: a parameter can be 
naturally small if setting it to zero leads to some enhanced symmetry in the system. This is usually the version of naturalness in which the concept of \textit{fine tuning} is understood~\cite{Barbieri}.
This principle of technical naturalness allows us to estimate the sizes of quantum corrections without carrying out explicit loop calculations. However, one should keep in mind that the actual loop results are usually more refined than solely applying technical naturalness, and can be utilized to optimize naturalness.

We will focus on a series of examples of  scalar QED in ($3+1$)-dimensional Aristotelian spacetime, with various dynamical critical exponents $z$. The scalar field in these toy models essentially plays the role of the Higgs in the Standard Model. We develop techniques that are useful for loop calculation in nonrelativistic gauge theories, and compute quantum corrections to the scalar propagator. The leading corrections to the scalar mass and speed of light are power law divergent. Such power law divergences and the associated naturalness in Aristotelian systems acquire new features, which we now summarize. 

First, in $3+1$ dimensions, the gauge coupling is dimensionful in theories around a Gaussian fixed point with $z>1$. Therefore, without turning on any marginal self-interaction term in the scalar sector, 
the theory is superrenormalizable. Relativistic superrenormalizable theories are usually not considered because power law divergences imply strong sensitivity to ultraviolet (UV) physics \cite{Zinn-Justin, schwartz}. However, in the Aristotelian case, the UV sensitivity is suppressed due to hierarchies in coupling constants supported by the polynomial shift symmetries \cite{nrn}.

Second, in theories with Aristotelian spacetime symmetries, there are no boosts that relate the temporal coordinate to spatial coordinates. In this case, there are naturally two different UV scales associated with time and space, respectively. This novelty opens up the question of how one should interpret power law divergences with respect to a two-parameter family of UV regulators. For logarithmic divergences, however, the evaluation of associated loop diagrams is independent of the UV regulation. Therefore, the beta functions in Aristotelian field theories around a given Gaussian fixed point are still well defined. We will discuss these important concepts in detail based on concrete examples in Section \ref{sec:pld} and Appendix \ref{app:logdiv}.

Another technical difficulty is the issue of gauge fixing. Aristotelian Yang-Mills theories were first introduced in \cite{qcYM}, and therein the temporal gauge was used. In \cite{Anselmi, Blas, pHL}, a gauge choice that manifestly respects the anisotropic spacetime scaling symmetry was introduced, which, as we will show in this paper, is analogous to the Lorenz gauge in the relativistic context. This Lorenz-type gauge is useful for us to explicitly check the gauge independence of any physical results. Alongside this novel type of gauge-fixing condition, we also provide a crosscheck in the more familiar Coulomb gauge. 
 
With all these technical developments in hand, we calculate the leading divergent quantum corrections to the scalar mass squared $m^2$ and its associated speed of light squared $c^2$ in an Aristotelian scalar QED with a $z=3$ Gaussian fixed point. We show that the suppression of power-law divergences in $m^2$, due to polynomial shift symmetries, can be further enhanced. This allows us to accommodate a sizable Yang-Mills coupling at low energies.If we let the scalar field and the $U(1)$ gauge field play the role of Higgs the $W$ and $Z$ gauge bosons in the Standard Model, then the enhancement of the smallness of $m$ allows us to maintain a hierarchy of 7 orders of magnitude between $m$ and the naturalness scale $M$ while keeping $\lambda_h \sim 1$, $y_f \lesssim 1$ and a gauge coupling of a realistic size, $g_{\text{YM}} \sim 0.1$. 

In Section~\ref{sec:review}, we review the mechanism proposed in \cite{nrn} and the way we suggest to improve it. In Sections~\ref{sec:rel_QED}, \ref{sec:QED_z2} and \ref{sec:QED_z3}, we compute the quantum corrections to the scalar propagator in relativistic scalar QED, and Aristotelian scalar QEDs with $z=2$ scaling and $z=3$ scaling, respectively. In particular, we mostly focus on the $z=2$ case in detail, which already exhibits the novelties of an Aristotelian quantum field theory (QFT) but with calculations which are not overly involved. In the $z=3$ case, the calculation proceeds in exactly the same manner, but is simply much more tedious and intricate. We will therefore present only the relevant results in the $z=3$ case. In Appendix \ref{app:logdiv} we study a QFT of a single scalar with constant shift symmetry in $2+1$ dimensions to illustrate the universality of beta functions in Aristotelian theories. Appendix \ref{app:bounds} discusses constraints on Lorentz violation parameters in the literature.

\section{Nonrelativistic Short-Distance Completions of the Higgs} \label{sec:review}

\subsection{Review of the model}

A nonrelativistic short-distance completion of a naturally light Higgs was proposed in \cite{nrn}. In the following, we summarize the main results of that work.

The theory contains a real massive scalar field $\phi (t, \by)$ which is assumed to be near a Gaussian fixed point in the UV, characterized by the dynamical critical exponent $z=3$. This is invariant under the time and space scaling
\be
    t \rightarrow b^z t, \qquad%
    \by \rightarrow b \, \by.
\ee
Operator dimensions are determined with respect to this $z=3$ scaling. The scalar field has the usual $\phi^4$ self-interaction, which is relevant in this case, but it may also have higher-derivative self-interactions with corresponding lower-dimension couplings. Reflection symmetry, $\phi \rightarrow - \phi$, and linear shift symmetry, $\delta \phi (t, \by) = b_i y^i$, are imposed on the scalar field, which limit the higher-derivative self-interactions to the one unique marginal or relevant interaction
\be 
    \CO = \p_i \phi \, \p_i \p_j \phi \, \p_j \p_k \phi \, \p_k \phi + \frac{1}{3} \p_i \phi \, \p_j \phi \, \p_k \phi \, \p_i \p_j \p_k \phi,
\ee 
The free and interaction parts of the action are thus taken to be
\begin{subequations}
\begin{align}
    S_2 &= \frac{1}{2} \int dt \, d^3 \by \biggl( \dot{\phi}^2 - \zeta_{3}^{2} \p^2 \p_i \phi \, \p^2 \p_i \phi - \zeta_{2}^{2} \p^2 \phi \, \p^2 \phi - c^2 \p_i \phi \, \p_i \phi - m^2 \phi^2 \biggr), \label{eq:S2} \\
    S_{\text{int}} &= - \frac{1}{2} \int dt \, d^3 \by \biggl( \lambda_3 \CO + \frac{1}{12} \lambda_0 \phi^4 \biggr). \label{eq:Sint}
\end{align}
\label{eq:scalaraction}
\end{subequations}
\!\!In the infrared (IR), all of the parameters of the theory are rescaled appropriately by powers of $c$ in order to set the coefficient of the two-derivative speed term $\p_i \phi \, \p_i \phi$ to 1,
\begin{subequations}
\begin{align}
    x^0 &= t, &%
    \bx &= \by / c, &%
    \Phi &= c^{3/2} \phi; \\
    \lambda_h &= \lambda_0 / c^3, &%
    \tilde{\zeta}_{3}^{2} &= \zeta_{3}^{2} / c^6, &%
    \tilde{\zeta}_{2}^{2} &= \zeta_{2}^{2} / c^4, &%
    \tilde{\lambda}_3 &= \lambda_3 / c^9,
\end{align}
\end{subequations}
in terms of which the action becomes
\be 
    S = \frac{1}{2} \int d^4 x \biggl[ \nabla_{\mu} \Phi \nabla^{\mu} \Phi - m^2 \Phi^2 - \frac{1}{12} \lambda_h \Phi^4 - \tilde{\zeta}_{3}^{2} ( \nabla_i \square \Phi )^2 - \tilde{\zeta}_{2}^{2} ( \square \Phi )^2 - \tilde{\lambda}_3 \tilde{\CO} \biggr],
\ee 
where $\tilde{\CO}$ is $\CO$ with $\p_i = \p / \p y^i$ and $\phi$ replaced respectively by $\nabla_i = \p / \p x^i$ and $\Phi$. Moreover, $\square \equiv \nabla_i \nabla_i$.

The pattern of symmetry breaking of polynomial shift symmetries in the UV action \eqref{eq:scalaraction} allows us to set up the technically natural hierarchy
\begin{align} \label{eq:hierarchy}
    \zeta_{3}^{2} &\sim 1, &%
    \lambda_3 &\sim \varepsilon_2, &%
    \zeta_{2}^{2} &\sim \varepsilon_2 M^{2/3}, &%
    c^2 &\sim \varepsilon_1 M^{4/3}, &%
    m^2 &\sim \varepsilon_0 M^2,
\end{align}
where $M$ is some high-energy naturalness scale, and
\be \label{eq:012}
    \varepsilon_0 \ll \varepsilon_1 \ll \varepsilon_2 \ll 1.
\ee 
The self-coupling $\lambda_0$ is also constrained to be small to not spoil the hierarchy between $m$ and $M$. The range of $\lambda_0$ is
\be
    \ell^{-1} \lambda_0 \lesssim \varepsilon_0 M^2.
\ee
We have included the one-loop suppression factor 
\be
\ell^{-1} = \frac{1}{16 \pi^2} \sim 10^{-2}.
\ee
From the perspective of a low-energy relativistic observer, the order of magnitude of the couplings are
\begin{subequations}
\begin{align}
    \lambda_h \lesssim \frac{\ell \varepsilon_0}{\varepsilon_1^{3/2}},
    \label{eq:boundlambdah}
\end{align}
\vspace{-5mm}
\begin{align}
    \tilde{\zeta}_3^2 &\sim \frac{\varepsilon_0^2}{\varepsilon_1^{3}} \frac{1}{m^4},&\!
    \tilde{\zeta}_2^2 &\sim \frac{\varepsilon_2 \varepsilon_0}{\varepsilon_1^2} \frac{1}{m^2},&\!
    \tilde{\lambda}_3 &\sim \frac{\varepsilon_2 \varepsilon_0^3}{\varepsilon_1^{9/2}} \frac{1}{m^6}.
\end{align}
\label{eq:lowrelvalues}
\end{subequations}

The Higgs sector is coupled to the fermions via the Yukawa interactions. In the IR, the range of the Yukawa couplings is
\be 
    y^{}_f \lesssim \frac{(\ell \varepsilon_{0})^{1/2}}{\varepsilon_{1}^{3/4}}.
    \label{eq:boundonYukawa}
\ee
The coupling to the gauge sector is achieved by introducing a $U(1)$ gauge field $a_{\mu}$ and covariantizing the partial derivatives acting on the scalar field $\phi$ via $\p_\mu \rightarrow D_\mu = \p_\mu + iea_\mu$, where $e$ denotes the gauge coupling. This leads to an IR Yang-Mills coupling
\be 
    g^{}_{\text{YM}} = \frac{e}{c^{1/2}}.
\ee 
Note that the couplings to gauge fields will generate divergent corrections to relevant terms (such as $\phi^6$ and $\phi^2 (\p_i \phi \, \p_i \phi)$) in \eqref{eq:scalaraction}, which means that such terms should be included in the theory. However, these quantum corrections are highly suppressed, which is consistent with technical naturalness.
Naively, the gauge coupling breaks any polynomial shift symmetries and one requires that
\be
    \ell^{-1} e^2 \lesssim \varepsilon_0 M^{2/3}.
\ee
Therefore,
\be
    g^{}_\text{YM} \lesssim \frac{(\ell \varepsilon_0)^{1/2}}{\varepsilon_1^{1/4}}.
    \label{eq:lowrelvaluesgYM}
\ee

We would like to see how large a hierarchy can be set up between the electroweak scale and a high-energy naturalness scale. We take the following numerical input from the Standard Model,
\be
    \lambda_h \sim 1,
        \qquad
    y_f \lesssim 1,
        \qquad
    g^{}_\text{YM} \sim 0.1.
    \label{eq:input}
\ee
To accommodate this set of values in \eqref{eq:boundlambdah}, \eqref{eq:boundonYukawa} and \eqref{eq:lowrelvaluesgYM}, we obtain the following conditions:
\be \label{eq:conds}
    1 \lesssim \frac{\ell \varepsilon_0}{\varepsilon_1^{3/2}},
        \qquad
    1 \lesssim \frac{(\ell \varepsilon_0)^{1/2}}{\varepsilon_1^{3/4}},  
        \qquad
    0.1 \lesssim \frac{(\ell \varepsilon_0)^{1/2}}{\varepsilon_1^{1/4}}.
\ee
Moreover, requiring that the Lorentz violations in \eqref{eq:lowrelvalues} be suppressed below the electroweak scale results in three additional conditions, 
\be
    \frac{\varepsilon_0^2}{\varepsilon_1^{3}} < 1,
        \qquad
    \frac{\varepsilon_2 \varepsilon_0}{\varepsilon_1^2} < 1,
        \qquad
    \frac{\varepsilon_2 \varepsilon_0^3}{\varepsilon_1^{9/2}} < 1.\label{eq:LV}
\ee
Note that the hierarchy between $M$ and $m$ is proportional to $\varepsilon_{0}^{-1/2}$. To maximize the hierarchy, we would like to minimize $\varepsilon_0$. As a consequence of the above conditions, we obtain,
\be
    \varepsilon_0 \sim 10^{-6},
        \qquad
    \varepsilon_1 \sim 10^{-4},
        \qquad
    \varepsilon_2 \sim 10^{-2},
\ee
with $10^{-6}$ the minimal value for $\varepsilon_0$ that we can achieve. In this case, we have
\be
    \frac{m}{M} \sim 10^{-3},
\ee
opening up only three orders of magnitude. This is in contrast to the capability of opening up 15 orders of magnitude in the ``10-20-30" model introduced in \cite{nrn}, in what is basically a gaugeless limit ($g^{\ }_\text{YM} \sim 10^{-10}$).

The situation gets worse when we apply this construction to the actual Standard Model. First, to accommodate the observed $W$ and $Z$ masses, one needs a Yang-Mills coupling of order $g^{\ }_\text{YM} \lesssim 0.65$ (instead of $g^{\ }_\text{YM} \sim 0.1$ in \eqref{eq:input}). Second, in a theory with $SU(3)$ gauge symmetry, one needs to sum over all 8 gluons, which introduces an extra factor of 8 in relevant Feynman diagrams. Taking into account both of these effects further reduces the available hierarchy and renders the naturalness scale to be about one order of magnitude higher than the electroweak scale.  

\subsection{Revisiting the gauge sector}

An interesting possibility for suppressing the quantum corrections to $m^2$ and $c^2$ from the gauge coupling is the following. Due to the presence of higher-derivative terms, covariantizing the partial derivatives in \eqref{eq:scalaraction} results in a number of different terms, which opens up the possibility of canceling the leading contributions to the Higgs mass among these terms. For example, covariantizing $(\p^2 \phi)^2$ results in three independent terms, namely,
\be \label{eq:D4phi}
    \overline{D^2 \phi} \, D^2 \phi,
        \qquad
    \overline{D_i D_j \phi} \, D_i D_j \phi,
        \qquad
    \overline{D_i D_j \phi} \, D_j D_i \phi.
\ee
We will refer to them as the $\zeta_2$ operators. Note that $\phi$ is now complex. We will, however, keep on using the same notation $\phi$ (and $\Phi$ for low-energy relativistic observers) for the complex scalar field. 

In the limit $\lambda_3 \rightarrow 0$, since both the gauge coupling and Yukawas have positive mass dimensions, the short-distance theory is superrenormalizable and the coupling constants only receive a classical renormalization group (RG) flow. Now, consider the corrections to $m^2$ coming from integrating out gauge fields. We organize these corrections in a perturbation series in $e$.

First, we consider just the contributions of the covariantized $\lr \p_i \p^2 \phi \rr^2$ operators, to which we refer as the ``$\zeta_3$ operators." Suppose that there exists a linear combination of the $\zeta_3$ operators such that the leading correction to $m^2$ (of order $\ell^{-1} \zeta_3^2 e^2 M^{4/3}$) vanishes. Then, such a condition will be preserved under the RG flow. This is due to the fact that the marginal couplings in front of the $\zeta_3$ operators only receive finite quantum corrections that can be removed by introducing finite counterterms. 

In the following discussion, we will assume that the order $\ell^{-1} \zeta_3^2 e^2 M^{4/3}$ correction to $m^2$ can be made to vanish. This assumption will be proven later in Section \ref{sec:QED_z3}. The next-to-leading order quantum corrections to $m^2$ from integrating out gauge fields now comes from two-loop diagrams of order $\ell^{-2} \zeta_3^2 e^4 M^\frac{2}{3} \sim \ell^{-2} e^4 M^{\frac{2}{3}}$, where we have used $\zeta_{3}^{2} \sim 1$.  

What about the contributions from the $\zeta_2$ operators in \eqref{eq:D4phi}? The leading contribution comes from a one-loop diagram of order $\ell^{-1} \zeta_2^2 e^2 M^\frac{2}{3}$. Finally, the leading contribution from the covariantization $D_i \phi \, D_i \phi$ of the $c^2$ term $\p_i \phi \, \p_i \phi$ is of order $\ell^{-1} c^2 e^2$, which is subleading to the contributions from $\zeta_2$ operators. Thus,
\be \label{eq:deltam2terms}
    \delta m^2 \sim \text{max} \left\{ \ell^{-2} e^4 M^{\frac{2}{3}} , \ell^{-1} \zeta_2^2 e^2 M^{\frac{2}{3}} \right\}.
\ee 
Depending on the size of $e^2$, one of these contributions to $m^2$ from $\zeta_3$ or $\zeta_2$ operators may dominate.\footnote{The reader might wonder how a two-loop effect could be comparable to or even dominant over a one-loop effect. This can happen if there are large hierarchies between coupling constants, as we have in our theory.}

We will estimate the size of $e^2$ using the quantum corrections to the $c^2$ term from integrating out the gauge fields. The leading-order correction here comes from the $\zeta_3$ operators and is of order $\ell^{-1} \zeta_3^2 e^2 M^{2/3} \sim \ell^{-1} e^2 M^{2/3}$. Therefore, in order that the condition $c^2 \sim \varepsilon_1 M^{4/3}$ be technically natural, we must have
\be 
    \ell^{-1} e^2 \lesssim \varepsilon_1 M^{2/3}.
    \label{eq:nrel}
\ee
The inequality \eqref{eq:nrel} turns \eqref{eq:deltam2terms} into $\delta m^2 \sim \varepsilon_1 \varepsilon_2 M^2$, where we have used the hierarchy $\varepsilon_1 \ll \varepsilon_2$ from \eqref{eq:012}. For the condition $m^2 \sim \varepsilon_0 M^2$ to be technically natural, we must have
\be
    \varepsilon_1 \varepsilon_2 \lesssim \varepsilon_0.
    \label{eq:120}
\ee

In this scenario, the conditions on the $\varepsilon$'s from the first two inequalities in \eqref{eq:conds} remain the same. However, importantly, the last condition in \eqref{eq:conds} is now modified to
\be
    10^{-1} = g^{}_\text{YM} = \frac{e}{c^\frac{1}{2}} \lesssim \frac{(\ell \varepsilon_1)^{1/2}}{\varepsilon_1^{1/4}} = \ell_{\phantom{1}}^{1/2} \varepsilon_1^{1/4}.
\ee
Given this new set of conditions, we can minimize $\varepsilon_0$, and thereby maximize the mass hierarchy, by choosing the following set of values for the $\varepsilon$'s:
\be
    \varepsilon_0 \sim 10^{-14}, 
        \qquad
    \varepsilon_1 \sim 10^{-8},
        \qquad
    \varepsilon_2 \sim 10^{-6}.
    \label{eq:1486}
\ee
Here, $\varepsilon_2$ is chosen such that \eqref{eq:120} is saturated. Hence,
\be
    \frac{m}{M} \sim 10^{-7},
    \label{eq:mM-7}
\ee
opening up 7 orders of magnitude between $m$ and $M$. 

Next, we turn on the $\lambda_3$ self-interaction in \eqref{eq:Sint}. Thanks to the nonrenormalization theorems proved in \cite{cmu, nrr}, Feynman diagrams that only contain vertices associated with the $\lambda_3$ self-interaction in $\phi$ do not generate any quantum corrections to $m^2$ or $c^2$. However, after covariantization, the $\lambda_3$ operator will give rise to nonzero quantum corrections to $m^2$ via diagrams that involve both $\phi$ and the gauge fields. This contribution to $m^2$ is bounded from above by $\ell^{-2} \lambda_3 e^2 M^{\frac{4}{3}}$. Applying \eqref{eq:nrel} and $\lambda_3 \sim \varepsilon_2$ in \eqref{eq:hierarchy}, we find that, given the choices of $\varepsilon$'s in \eqref{eq:1486}, the hierarchy in \eqref{eq:mM-7} is preserved.

It may seem surprising that we are free to choose the marginal coefficients in front of the $\zeta_3$ operators such that the leading divergence in $m^2$ vanishes, without resorting to any extra symmetries. However, it turns out that the linear shift symmetry is sufficient to do the job. We can see this as follows. Usually, a marginal self-coupling such as $\lambda_3$ is naturally of order one. Turning on such a marginal coupling would cause the coefficients of the $\zeta_3$ operators to run strongly, such that the initial choice of those $\zeta_3$ operators is completely spoiled under renormalization. In our case, however, the self-coupling $\lambda_3$ (and its quantum corrections to other parameters) is protected to be very small due to the linear shift symmetry. In this sense, it is still the polynomial shift symmetry that protects the smallness of $m^2$, and this scenario is an example of how naturalness can be ``optimized."

This method for opening up the hierarchy while keeping the value of $g^{\ }_\text{YM}\sim 0.1$ requires us to be able to set the couplings of the theory so as to set the order $\ell^{-1} e^2 M^\frac{4}{3}$ correction to $m^2$ to zero, in the absence of marginal interactions. In the following sections, we calculate this order $e^2$ correction to $m^2$ in the relativistic case, in nonrelativistic scalar QEDs with $z=2$ and $z=3$ scaling and indeed show that in the two latter cases, this correction can be set to zero. 

If we want to further increase the hierarchy between $m$ and $M$, it appears that we will need to set the leading correction to $c^2$ to zero. If this were the case, then we would be able to set $\varepsilon_0 \sim 10^{-18}$ and hence open up a hierarchy of 9 orders of magnitude. However, we will demonstrate explicitly that this is not possible. This claim may sound surprising, since naively there is only a single quadratically divergent (measured in momentum scale) correction to $c^2$. Nevertheless, in the nonrelativistic case, as we will demonstrate in the paper, there exist distinct contributions that are all quadratically divergent by power counting. We will show that these cannot be made to cancel among themselves.

\subsection{Low-energy suppression of Lorentz violation}

Finally, we would like to comment on the Lorentz symmetry restoration in the IR. We have required \eqref{eq:LV} such that all Lorentz violating operators are suppressed. This is, however, not sufficient for the Lorentz symmetry to be recovered. In addition, we will have to require that the speeds of light of different species of particles (collectively denoted by $c_i^2$ in the following) be the same. In the above construction, after turning on the universal gauge coupling $e$, from the UV perspective, all $c_{i}^{2}$ receive a quantum correction of the size $\ell^{-1} e^2 M^{2/3}$. This is the largest quantum correction to $c_i^2$. It is therefore technically natural to take all $c_i^2$ to be of the same order with 
\be
	c_i^2 = \CO (\ell^{-1} e^2 M^{2/3}) = \CO ( \varepsilon^{}_1 M^{4/3}).
\ee
Hence, it is also technically natural to take the initial condition that all $c_i^2$ are equal.\footnote{Technical naturalness does not explain why parameters take on some specific set of values. For example, we try to use this principle to explain why the Higgs mass \textit{can} be of order 0.1-1 TeV, not why it actually \textit{is} of this order or, even more to the point, why it has the specific value 125 GeV. Similarly, this principle allows the $c^2$ parameters of all species to be of order $\varepsilon_1 M^{4/3}$, but explaining why the speeds are all the same is beyond the scope of technical naturalness and requires detailed knowledge of the fundamental theory from which this theory descends.}


Does the matching condition among $c_i^2$ survive in the IR? This question requires us to look into the logarithmic divergences and estimate the beta functions of various $c_i^2$. By dimensional analysis, the leading logarithmic divergence is proportional to $e^4$, and hence the physical values of $c_i^2$ take the following form: 
\begin{align}
\label{eq:corrci}
    c_i^2 = c_{0}^2 + \mathcal{C}_i \ell^{-2} \, e^4 \log M + \text{higher order terms},
\end{align}
where $\mathcal{C}_i$ is an order one constant coefficient. We have taken all $c_i^2$ to have the same initial value $c_0^2 = \mathcal{O} (\varepsilon_1 M^{4/3})$. From \eqref{eq:corrci}, we obtain the anomalous dimension for $c_i^2$,
\begin{align}
    \gamma_{c_i^2} & = \mathcal{C}_i \ell^{-2} \frac{e^4}{c_0^2} + \text{higher order terms} = \mathcal{O} (\varepsilon_1 M^\frac{4}{3}).
    \label{eq:gammaci2}
\end{align}
To derive the associated beta function, we first define a dimensionless coupling via
\begin{align}
    c_i^2 = \rho_i M^\frac{4}{3}.
\end{align}
Here, $\rho_i$ characterizes the importance of the $c_i^2$ operator at the scale $M$. The beta function associated with $c_i^2$ is defined as follows:
\begin{align}
   \beta_{i} & \equiv \frac{d \rho_i}{d \log M}
   = \left ( - \frac{4}{3} + \gamma_{c_i^2} \right ) \rho_i. 
\end{align}
The first term $- 4/3$ comes from the classical dimension of $c_i^2$, which is common for all $c_i^2$; the coefficient $\mathcal{C}_i$ in $\gamma_{c_i^2}$ may vary for different $c_i^2$. However, this species-dependent part in the beta function is suppressed by a factor of $\varepsilon_1 \ll 1$, as indicated in \eqref{eq:gammaci2}. Therefore, the matching condition among $c_i^2$ is preserved in the IR up to hierarchically small corrections of order $\varepsilon_1$.\footnote{More precisely, the corrections are of order $\varepsilon_1 \log (M / m)\sim 7 \epsilon_1$.}

\section{Relativistic Scalar QED} \label{sec:rel_QED}

We give a short review of the relativistic scalar QED in $3+1$ dimensions. We will use the mostly negative signature, with the metric $\eta_{\mu\nu} = (+, -, -, -)$. The gauge field $A_\mu$, $\mu = 0, 1, \ldots, 3$ is a one-form on spacetime. The $U(1)$ gauge transformation acts on $A_\mu$ in the usual way,
\be
    \delta_\epsilon A_\mu = \frac{1}{g} \nabla_\mu \epsilon. 
\ee
We couple a complex scalar field $\Phi$ to the gauge field in a way that preserves the $U(1)$ gauge symmetry, which requires that $\Phi$ transform as
\be
    \Phi \rightarrow e^{- i \epsilon} \Phi,
\ee
where $g$ is the gauge coupling. The invariant action is
\be
\label{eq:z1action}
    S = \int d^4 x \lc - \frac{1}{4} F_{\mu\nu} F^{\mu\nu} + \oln{\mathcal{D}_\mu \Phi} \, \mathcal{D}^\mu \Phi - m^2 \oln{\Phi} \Phi - \lambda \bigl(\oln{\Phi} \Phi\bigr)^2 \rc,
\ee
where the covariant derivative $D_\mu = \nabla_\mu + i g A_\mu$, with $g$ the gauge coupling and $F_{\mu \nu}$ is the antisymmetric field strength tensor $F_{\mu \nu} = \nabla_{\mu} A_{\nu} - \nabla_{\nu} A_{\mu}$. We would like to calculate the quadratically divergent correction to the scalar mass squared due to the scalar-gauge interaction by using the Coulomb gauge and the Lorenz gauge. Working in different gauges provides us with a powerful crosscheck of the results.

\subsection{One-loop correction in Coulomb gauge}

We calculate the one-loop correction to the mass squared of the scalar field $\Phi$ in the Coulomb gauge. We start by expanding the gauge sector of the action \eqref{eq:z1action} in terms of its components $A_0$ and $A_i$,
\bea
S_\text{A} &=& - \frac{1}{4} \int d^4 x \, F_{\mu\nu} F^{\mu\nu} \notag \\
    &=& \frac{1}{2} \int d^4 x \left\{ \dot{A}_i \dot{A}_i - 2\dot{A}_i \nabla_i A_0 +  \nabla_i A_0 \nabla_i A_0 -  \nabla_i A_j \nabla_i A_j +  \nabla_i A_j \nabla_j A_i \right\}.
\eea
In Coulomb gauge, $A_0$ is not dynamical and needs to be integrated out in the path integral, which essentially induces an instantaneous Coulomb interaction between charge densities. Due to the presence of nonlinear terms, it is more convenient to keep $A_0$ in the path integral and integrate it out as internal legs in Feynman diagrams. 

Both the Coulomb gauge condition ($\nabla_i A_i = 0$) and Gauss constraint ($\nabla_i E_i = 0$, where $E_i = \nabla_0 A_i - \nabla_i A_0$ is the electric field) are second class, which means that the commutation relations should be given by the associated Dirac brackets (up to a prefactor $i$). From the appropriately defined Dirac brackets, we derive the Feynman rules for the propagators of the gauge fields,
\be
\begin{minipage}{5cm}
\begin{tikzpicture}
\draw[decorate,decoration=snake] (0,0) -- (2,0) ;
\small \node [left] at (0,0) {$A^\mu$};
\node [right] at (2,0) {$A^\nu$};
\node [right] at (0.8,0.3) {$k$};
\node [right] at (0.8,-0.3) {$\phantom{k}$};
\draw [->] [thick]   (1.09,-0.0) -- (1.10,0.01);
\end{tikzpicture}
\end{minipage}
\Delta^{\mu\nu} (k) = \begin{pmatrix}
            \frac{i}{|\bk|^2} & 0 \\
            0 & \frac{i}{k^2 + i \epsilon}\left( \delta^{ij} - \frac{k^i k^j}{|\bk|^2}\right)
        \end{pmatrix}.
\label{eq:reCgp}
\ee
The timelike component (\emph{i.e.}, the $\Delta^{00}(k)$ propagator) does not have a physical pole. Since $A_0$ can never be put on shell, it will only appear in internal legs of Feynman diagrams. 

On the other hand, the scalar part of the action \eqref{eq:z1action} is
\bea
S_{\Phi} &=& \!\! \int \! d^4 x \, \Bigl [ \oln{\mathcal{D}_{\!\mu} \Phi} \, \CD^{\mu} \Phi - m^2 \oln{\Phi} \Phi - \lambda (\oln{\Phi} \Phi)^2 \Bigr ] \nonumber \\
&=& \! \int \!\! d^4 x \, \Bigl [ \nabla_{\!\mu} \oln{\Phi} \, \nabla^\mu \Phi - i g A^\mu \bigl(\oln{\Phi} \, \nabla_{\!\mu} {\Phi} - \Phi \, \nabla_{\!\mu} \oln{\Phi} \bigr) + g^2 A_\mu A^\mu \oln{\Phi} \Phi  - m^2 \oln{\Phi} \Phi - \lambda (\oln{\Phi} \Phi)^2 \Bigr ].
\eea
From the scalar action $S_{\Phi}$, we derive the Feynman rules for the scalar field propagator 
\be \label{eq:scalarprop}
\begin{minipage}{5cm}
\begin{tikzpicture}
\draw [->] (0,0) -- (1,0);
\draw [->] [thick] (1,0) -- (1.01,0);
\draw (1,0) -- (2,0);
\node [white] at (1,-0.3) {$k$};
\node at (1,0.3) {$k$};
\small \node [left] at (0,0) {$\oln{\Phi}$};
\node [right] at (2,0) {$\Phi$};
\end{tikzpicture}
\end{minipage}
\hspace{-1cm}
\Delta(k) =  \frac{i}{k^2 - m^2 + i \epsilon},
\ee
and the gauge interaction vertices
\begin{subequations}
\begin{align}
&
\begin{minipage}{6cm}
\begin{tikzpicture}[scale=0.8]
\draw[decorate,decoration=snake] (0,-0.5) -- (0,0.5) ;
\draw[->] (-1,-1.5) -- (-0.5,-1) ;
\draw[->][thick] (-0.5,-1) -- (-0.49,-0.99) ;
\draw (-0.5,-1) -- (0,-0.5) ;
\draw[->] (1,-1.5) -- (0.5,-1) ;
\draw[->][thick] (0.5,-1) -- (0.49,-0.99) ;
\draw (0.5,-1) -- (0,-0.5) ;
\draw[->][thick] (-0.01,-0.01) -- (.0,-0.02) ;
\small \node [below] at (-1,-1.5) {$\oln{\Phi}\left(\qk_1\right)$};
\node [below] at (1,-1.5) {$\Phi\left(\qk_2\right)$};
\small \node [above] at (0,0.5) {$A_{\mu}(\qq)$};
\end{tikzpicture}
\end{minipage}
\hspace{-3.2cm}
V_\mu (\qq, k_1, k_2) = - i g \lr\qku - \qkd\rr_\mu, \label{eq:tpt} \\[10pt]
&
\begin{minipage}{6cm}
\begin{tikzpicture}[scale=0.8]
\draw[decorate,decoration=snake] (0,-0.5) -- (1,0.5) ;
\draw[->][thick] (0.46,-0.01) -- (0.44,-0.01) ;
\draw[decorate,decoration=snake] (0,-0.5) -- (-1,0.5) ;
\draw[->][thick] (-0.5,-.07) -- (-0.50,-0.08) ;
\draw[->] (-1,-1.5) -- (-0.5,-1) ;
\draw[->][thick] (-0.5,-1) -- (-0.49,-0.99) ;
\draw (-0.5,-1) -- (0,-0.5) ;
\draw[->] (1,-1.5) -- (0.5,-1) ;
\draw[->][thick] (0.5,-1) -- (0.49,-0.99) ;
\draw (0.5,-1) -- (0,-0.5) ;
\small \node [below] at (-1,-1.5) {$\oln{\Phi}\left(\qku\right)$};
\node [below] at (1,-1.5) {$\Phi\left(\qkd\right)$};
\small \node [above] at (1,0.5) {$A_{\nu}(\qp)$};
\small \node [above] at (-1,0.5) {$A_{\mu}(\qq)$};
\end{tikzpicture}
\end{minipage}
\hspace{-2.6cm}
V_{\mu\nu} (\qp, \qq, \qku, \qkd) = 2 i g^2 \eta_{\mu \nu}. \label{eq:fpt}
\end{align}
\end{subequations}
Since we are not interested in the IR behavior, we will simply set $m$ to zero in the following calculation; the integrals are understood to be regulated in the IR, however.

At one-loop order, the contributions to the quadratic divergence of the scalar mass come from the following Feynman diagrams. As a convention, we define frequencies and momenta in pairs, such as $k = (\omega, \bk)$, $p = (\eta, \bp)$ and $q = (\nu, \bq)$. There are the ``cog" diagram,
\begin{align}
\begin{minipage}{7cm}
\begin{tikzpicture}
\draw  [decorate,decoration=snake] (0,0.77) circle (0.77);
\draw  [->][thick]  (82:1.63)  -- + (20:-0.01);
\draw   (-1.7,0) -- (1.75,0);
\draw [->] [thick]   (-1.21,0) -- (-1.2,0);
\draw [<-]  [thick] (1.3,0) -- (1.31,0);
\node [left] at (-1.2,-0.4) {$\overline{\Phi} (\qk)$};
\node [right] at (1.2,-0.4) {$\Phi (-\qk)$};
\node [above][ultra thick] at (0.1,1.65) {$\qq = (\nu, \bq)$};
\end{tikzpicture}
\end{minipage}
\hspace{-2cm}
&= \frac{1}{2} \int \frac{d^4 q}{(2\pi)^4} \, V_{\mu\nu} (\qq, -\qq, \qk, -\qk) \, \Delta^{\mu\nu} (q) \notag \\[-25pt]
&= g^2 \int \frac{d^4q}{(2\pi)^4} \lr \frac{2}{\qq^2 + i\epsilon} - \frac{1}{|\bqq|^2} \rr,
\intertext{and the ``sunset" diagram,}
\begin{minipage}{7cm}
\begin{tikzpicture}
\begin{scope}
\clip (-1,0) rectangle (1,1);
\draw  [decorate,decoration=snake] (0,0) circle (0.75);
\end{scope}
\draw  [->][thick]  (0.0,0.82)  -- (-.01,.83) ;
\draw  [<-][thick]  (250:0)  -- + (0:0.01) ;
\draw  (-1.7,0) -- (1.75,0);
\draw [->]  [thick] (-1.21,0) -- (-1.2,0);
\draw [<-]  [thick] (1.3,0) -- (1.31,0);
\node [left] at (-1.2,-0.4) {$\overline{\Phi} (\qk)$};
\node [right] at (1.2,-0.4) {$\Phi (-\qk)$};
\node [above] at (0.1,0.9) {$\qq = (\qn, \bqq)$};
\node [below] at (0.1,-0.1) {$\qp$};
\end{tikzpicture}
\end{minipage}
\hspace{-2cm}
&= \int \frac{d^4 q}{(2\pi)^4} \, V_{\mu}(q,k,p) \, \Delta^{\mu \nu}(q) \, V_{\nu} (-q,-p,-k) \, \Delta(p) \notag \\[-15pt]
&= g^2 \int \frac{d^4 q}{(2\pi)^4} \frac{\nu^2}{|\mathbf{q}|^2} \frac{1}{q^2 + i\epsilon} + \cdots,
\end{align}
where ``$\cdots$" contains subleading divergences. We have set the external momenta to zero in order to extract the quantum corrections to the nonderivative mass term.

The total leading divergence from both diagrams is
\be
 i \Gamma_A
=  3 g^2 \int \frac{d^4 \qq}{(2\pi)^4} \frac{1}{q^2}. 
\label{eq:GammaA}
\ee
As expected, Lorentz symmetry has been recovered. Moreover, this result is nonsingular despite the fact that $\Delta^{00} (x)$ has no physical pole. In the sharp cutoff regularization scheme, this integral is regulated in the UV by introducing a cutoff $M$ for the four-momentum $q$, with $0 < |q| < M$, such that
\be
    \Gamma_A = - \frac{3}{16 \pi^2}  g^2 M^2.
\ee
The $\lambda$ term also contributes a quadratic divergence to $m^2$, which is given by
\be
    \Gamma_\Phi = - \frac{1}{16 \pi^2}  \lambda M^2.
\ee
Therefore, it is technically natural to take
\be
    m^2 \sim \text{max} \{ g^2, \lambda \} \, M^2.
\ee
This leads to the usual statement of the naturalness problem of a massive scalar: for typical values of $g$ and $\lambda$ not much smaller than 1, the Higgs mass $m$ is naturally of order $M$.

\subsection{One-loop correction in Lorenz gauge}

Next, we repeat the above calculation by applying the Faddeev-Popov method. We apply the Lorenz gauge by taking the following gauge-fixing functional:
\be 
    f [A^\mu] = \nabla_\mu A^\mu.
\ee
The gauge-fixed action is the sum of the original action $S = S_A + S_\Phi$ and a gauge fixing term $S_\text{g.f.}$, which explicitly breaks gauge invariance. The gauge fixing term $S_\text{g.f.}$ is
\be \label{eq:relgf}
    S_\text{g.f.} = - \frac{1}{2\xi} \int d^4 x \left(\nabla_{\mu} A^{\mu}\right)^2
\ee 
One also introduces ghost fields in the standard way. For the $U(1)$ gauge theory, the ghost fields do not contribute to the Feynman diagrams. The choice of $\xi$ determines the choice of gauge. For example, the Feynman--'t Hooft gauge is given by $\xi = 1$, and $\xi \rightarrow 0$ gives the Landau gauge. 
The Lorenz gauge propagator can be written as 
\be
\begin{minipage}{5cm}
\begin{tikzpicture}
\draw[decorate,decoration=snake] (0,0) -- (2,0) ;
\small \node [left] at (0,0) {$A^\mu$};
\node [right] at (2,0) {$A^\nu$};
\node [right] at (0.8,0.3) {$k$};
\node [right] at (0.8,-0.3) {$\phantom{k}$};
\draw [->] [thick]   (1.09,-0.0) -- (1.10,0.01);
\end{tikzpicture}
\end{minipage}
\Delta^{\mu\nu}
=
\frac{-i}{k^2 + i \epsilon} \left [ \eta^{\mu\nu} - (1-\xi) \frac{k^\mu k^\nu}{k^2 + i \epsilon} \right ].
\label{eq:Rxirel}
\ee

The Feynman rules for the gauge interactions have been given in \eqref{eq:tpt} and \eqref{eq:fpt}. 
At the one-loop order, we have the cog diagram
\begin{align}
\begin{minipage}{7cm}
\begin{tikzpicture}
\draw  [decorate,decoration=snake] (0,0.77) circle (0.77);
\draw  [->][thick]  (82:1.63)  -- + (20:-0.01);
\draw   (-1.7,0) -- (1.75,0);
\draw [->] [thick]   (-1.21,0) -- (-1.2,0);
\draw [<-]  [thick] (1.3,0) -- (1.31,0);
\node [left] at (-1.2,-0.4) {$\overline{\Phi} (\qk)$};
\node [right] at (1.2,-0.4) {$\Phi (-\qk)$};
\node [above][ultra thick] at (0.1,1.65) {$\qq$};
\end{tikzpicture}
\end{minipage}
\hspace{-2cm}
&= \frac{1}{2} \int \frac{d^4 q}{(2\pi)^4} \, V_{\mu\nu} (\qq, -\qq, \qk, -\qk) \, \Delta^{\mu\nu} (q) \notag \\[-25pt]
&= g^2 \left(3 + \xi \right) \int \frac{d^4 \qq}{(2\pi)^4} \frac{1}{\qq^2},\\
\intertext{and the ``sunset" diagram}
\begin{minipage}{7cm}
\begin{tikzpicture}
\begin{scope}
\clip (-1,0) rectangle (1,1);
\draw  [decorate,decoration=snake] (0,0) circle (0.75);
\end{scope}
\draw  [->][thick]  (0.0,0.82)  -- (-.01,.83) ;
\draw  [<-][thick]  (250:0)  -- + (0:0.01) ;
\draw  (-1.7,0) -- (1.75,0);
\draw [->]  [thick] (-1.21,0) -- (-1.2,0);
\draw [<-]  [thick] (1.3,0) -- (1.31,0);
\node [left] at (-1.2,-0.4) {$\overline{\Phi} (\qk)$};
\node [right] at (1.2,-0.4) {$\Phi (-\qk)$};
\node [above] at (0.1,0.9) {$\qq$};
\node [below] at (0.1,-0.1) {$\qp$};
\end{tikzpicture}
\end{minipage}
\hspace{-2cm}
&= \int \frac{d^4 q}{(2\pi)^4} \, V_{\mu}(q,k,p) \, \Delta^{\mu \nu}(q) \, V_{\nu} (-q,-p,-k) \, \Delta(p) \notag \\[-15pt]
&= - g^2 \xi \int \frac{d^4 \qq}{(2\pi)^4}  \frac{1}{\qq^2} + \cdots. 
\end{align}

The sum over these contributions recovers the result in \eqref{eq:GammaA}. As expected, the gauge parameter $\xi$ drops out of the sum.

\section{Aristotelian Scalar QED with \texorpdfstring{$z=2$}{z=2} Scaling} \label{sec:QED_z2}

In this section, we consider a slightly more complicated theory in $3+1$ dimensions around a $z=2$ Gaussian fixed point. This 
will help us build up useful intuition before moving on to the even more complicated $z=3$ scenario. 

To distinguish from the relativistic case, we denote the gauge fields 
by $a_\mu = (a_0, a_i)$, $i = 1,\ldots,3$ in this section and the next. The $U(1)$ gauge transformations are the same as in the relativistic case,
\be
    \delta a_0 = \frac{1}{e}\dot{\epsilon},
        \qquad
    \delta a_i = \frac{1}{e} \p_i \epsilon.
\ee
Here, $\epsilon$ is a dimensionless parameter. Around a Gaussian fixed point with a dynamical exponent $z$, the engineering dimensions for time and space coordinates $t$ and $\by$ measured in energy are
\be
    [t] = -1,
        \qquad
    [\by] = - \frac{1}{z}.
\ee

Around a $z=2$ Gaussian fixed point, the gauge action is
\begin{align} \label{eq:Sa}
    S_a = \int dt \, d^3 \by \Bigl ( \frac{1}{2} E_i E_i - \frac{1}{4} \zeta^2_{2,a} \p_k F_{ij} \p_k F_{ij} - \frac{1}{4} c_a^2 F_{ij} F_{ij} \Bigr ).
\end{align}
where $E_i = F_{0i}$. In this normalization, the engineering dimensions for $a_0$ and $a_i$ are
\be
    [a_0] = \frac{3}{4},
        \qquad
    [a_i] = \frac{1}{4}.
\ee

Next, we construct the scalar action in the limit of zero gauge coupling $e \rightarrow 0$. The engineering dimension of the scalar field $\phi$ is 
\be
    [\phi] = \frac{1}{4}.
\ee
In addition to the Aristotelian spacetime symmetries, we also impose time reversal symmetry $\CT: t \rightarrow - t$, which acts on the field trivially, and reflection symmetry in the field $\phi \rightarrow - \phi$. We consider the following scalar action that preserves the constant shift symmetry:
\be
    S_{\phi, \, e = 0} = \int dt \, d^3 \by \Bigl( \p_0 \oln{\phi} \, \p_0 \phi - \zeta_2^2 \p^2 \oln{\phi} \, \p^2 \phi - c^2 \p_i \oln{\phi} \, \p_i \phi \Bigr),
    \label{eq:Sphie0}
\ee
with the engineering dimensions,
\bea 
[\zeta_2^2]=0, \qquad [c^2] = 1.
\eea 
With applications to the Higgs in mind, we also turn on the nonderivative terms in $S_{\phi, \, e=0}$ in \eqref{eq:Sphie0},
\be
    - \int dt \, d^3\by \, \Bigl [ m^2 \oln{\phi} \phi + \lambda_0 \bigl( \oln{\phi} \phi \bigr)^2 \Bigr ],
\ee
which break the constant shift symmetry in the softest possible way. 

Let us then define the covariant derivatives $D_0$ and $D_i$, $i = 1,2, 3$, to be
\be
    D_0 = \p_0 + i e a_0,
        \qquad
    D_i = \p_i + i e a_i.
\ee
There are three different ways of covariantizing $\p^2 \oln{\phi} \, \p^2 \phi$,
\be \label{eq:D4phi2}
    \oln{D^2 \phi} \, D^2 \phi, 
        \qquad 
    \oln{D_i D_j \phi} \, D_i D_j \phi, 
        \qquad  
    \oln{D_i D_j \phi} \, D_j D_i\phi.
\ee
It is useful to note the following identities:
\bea
    \int dt \, d^3 \by \, \oln{D_i D_j \phi} \, D_j D_i \phi &=& \int dt \, d^3 \by \, \lr \oln{D^2 \phi} \, D^2 \phi + i e F_{ij} \oln{D_i \phi} \, D_j \phi \rr, \\[2pt]
    \int dt \, d^3 \by \, \oln{D_i D_j \phi} \, D_i D_j \phi &=& \int dt \, d^3 \by \, \lr \oln{D^2 \phi} \, D^2 \phi + i e F_{ij} \oln{D_i \phi} \, D_j \phi + \tfrac{1}{2} e^2 F_{ij} F_{ij} \oln{\phi} \phi \rr.
\eea
Hence, we find a more convenient basis that is equivalent (up to total derivatives) to \eqref{eq:D4phi2},
\be
    \oln{D^2 \phi} \, D^2 \phi,
        \qquad
    i e F_{ij} \oln{D_i \phi} \, D_j \phi,
        \qquad
    \frac{1}{2} e^2 F_{ij} F_{ij} \oln{\phi} \phi.
    \label{eq:diffbasis}
\ee
In this basis, the Feynman rules are easier to deal with. Finally, the covariantized scalar action is
\begin{align}
S_{\phi} = \int dt \, d^3 \textbf{y} \Bigl [ \oln{D_0 \phi} \, D_0 \phi - \zeta_2^2 \oln{D^2 \phi} \, D^2 \phi & - c^2  \oln{D_i \phi} \, D_i \phi - m^2 \oln{\phi} \phi - \lambda_0 (\oln{\phi} \phi)^2 \notag \\ 
& - i e \eta_1 F_{ij} \oln{D_i \phi} \, D_j \phi - \frac{e^2}{2} \eta_2 F_{ij} F_{ij} |\phi|^2 \Bigr ].
\end{align}
From this scalar action $S_{\phi}$, we derive the Feynman rule for the scalar field propagator 
\begin{align} \label{eq:scalarpropz2}
\begin{minipage}{5cm}
\begin{tikzpicture}
\draw [->] (0,0) -- (1,0);
\draw [->] [thick] (1,0) -- (1.01,0);
\draw (1,0) -- (2,0);
\node [white] at (1,-0.3) {$k$};
\node at (1,0.3) {$k$};
\small \node [left] at (0,0) {$\oln{\Phi}$};
\node [right] at (2,0) {$\Phi$};
\end{tikzpicture}
\end{minipage} 
\Delta(k) =  \frac{i}{\omega^2 - \zeta_2^2 |\bk|^4 - c^2 |\bk|^2 - m^2  + i \epsilon}.
\end{align}

The theory is given by the action 
\be
    S = S_a + S_\phi.
\ee
The coupling constants $\zeta^2_{2,a}$ and $\zeta_2^2$ do not receive divergent corrections. We will set $\zeta^2_{2, a} = \zeta_2^2 = 1$. Moreover, we will tune $c^2_a$ and $c^2$ such that $c_a^2 = c^2$; as argued in Section \ref{sec:review}, there is no fine tuning required. Furthermore, this allows for the emergence of Lorentz symmetry at low energies and not just $z=1$ scaling.

\subsection{The gauge propagator}

First, we focus on the gauge sector described by the action \eqref{eq:Sa}. Instead of directly quantizing the theory with the full action $S = S_a + S_\phi$, let us consider a simpler case where the gauge field $a_\mu$ is coupled linearly to a nondynamical external source $J^\mu = (-\rho, J^i)$. We compute the partition function,
\be
    \CZ [J] = \int \CD a_\mu \, \exp \! \ls i S_a + i \int dt \, d^3\by \, (-\rho a_0 + J^i a_i) \rs.
    \label{eq:paflin}
\ee
Requiring that the source term be gauge invariant forces the external source $J^\mu$ to be a conserved current, \emph{i.e.},
\be
    \dot{\rho} - \p_i J^i = 0.
    \label{eq:clc}
\ee
In components, the action is
\begin{align}
S_{a} &= \int dt \, d^3 \textbf{y} \left(\frac{1}{2}E_i E_i - \frac{1}{4}\p_k F_{ij} \p_k F_{ij}  - \frac{c_a^2}{4} F_{ij} F_{ij} \right) \nonumber \\
&= \frac{1}{2} \int dt \, d^3 \textbf{y} \left[ \dot{a}_i \dot{a}_i - 2 \dot{a}_i \p_i a_0 +  \left(\p_i a_0\right)^2  +  a_i (- \p^2 + c_a^2) (\delta_{ij} \p^2 - \p_i \p_j) a_j \right]. 
\end{align}
Here, $\p^2 \equiv \p_i \p_i$. We have kept the $c_a^2$ terms for completeness of the formal discussion. It is convenient to define
\be
    \bK \equiv \bk \sqrt{|\bk|^2 + c_a^2}.
\ee
We also define $K_\mu \equiv (\omega, \bK)$ and $K^2 \equiv \omega^2 - |\bK|^2$.

The quantization in Coulomb gauge (with the gauge condition $\p_i a_i = 0$) proceeds almost identically as for the relativistic case, except that the dispersion relation for $a_i$ is modified. In Coulomb gauge, the photon propagator is 
\be
\begin{minipage}{5cm}
\begin{tikzpicture}
\draw[decorate,decoration=snake] (0,0) -- (2,0) ;
\small \node [left] at (0,0) {$a^\mu$};
\node [right] at (2,0) {$a^\nu$};
\node [right] at (0.8,0.3) {$k$};
\node [right] at (0.8,-0.3) {$\phantom{k}$};
\draw [->] [thick]   (1.09,-0.0) -- (1.10,0.01);
\end{tikzpicture}
\end{minipage}
\Delta^{\mu\nu}_\text{Coul} (k) = \begin{pmatrix}
            \frac{i}{|\bk|^2} & 0 \\
            0 & \frac{i}{K^2 + i \epsilon}\left( \delta_{ij} - \frac{k_i k_j}{|\bk|^2}\right)
        \end{pmatrix}.
    \label{eq:coulombz2}
\ee

In this simple case, it is more convenient to directly eliminate the nondynamical field component $a_0$ by enforcing its equation of motion,
\be
    \rho = - \p^2 a_0. 
\ee
After eliminating $a_0$ and integrating out $a_i$ in the partition function, we obtain 
\begin{align}
    \CZ [J] & = \exp \ls - \frac{i}{2} \int dt \, V_\text{Coul} (t) \rs \exp \ls - \frac{i}{2} \int d^4 y \,  d^4 y' \, J_i(y) \Delta^{ij}_\text{Coul}(y-y') J_j(y') \rs \notag \\[2pt]
        & = \exp \ls - \frac{i}{2} \int d^4 y \, d^4 y' \, J_\mu(y) \Delta^{\mu\nu}_{\text{eff}}(y-y') J_\nu(y') \rs,
\label{eq:ZJDeff}
\end{align}
where $V_\text{Coul} (t)$ denotes the Coulomb potential energy as in relativistic QED,
\be
    V_\text{Coul} (t) = \int d^3 \by \, d^3 \by' \frac{\rho (t, \by) \, \rho (t, \by')}{4 \pi |\by - \by'|},
\ee
and in the frequency-momentum space we have
\be
    \Delta^{\mu \nu}_\text{eff} (k) = \frac{i}{K^2 + i \varepsilon}
        \begin{pmatrix}
            - (|\bk|^2 + c_a^2) & \,\, 0 \, \\[2pt]
            0 & \,\, \delta_{ij} \,
        \end{pmatrix}.
\label{eq:Deff}
\ee
We have used the conservation law of current \eqref{eq:clc} to derive \eqref{eq:ZJDeff}. 

In the $z=2$ scalar QED that we are considering in this section, there are nonlinear terms in $a_0$ and the scalar field is also dynamical. It is more convenient to work directly with the singular propagator in \eqref{eq:coulombz2} for loop calculations.

One can set the first entry $- (|\bk|^2 + c_a^2)$ of the matrix in \eqref{eq:Deff} to $-1$ by taking a nonlocal field redefinition, 
\be
    a_0 \rightarrow \sqrt{- \p^2 + c_a^2} \, a_0,
    \label{eq:fda0}
\ee
resulting in 
\be \label{eq:coulombpropeff}
    \Delta^{\mu\nu}_\text{eff} = - \frac{i \eta^{\mu\nu}}{K^2 + i \varepsilon}.
\ee
This is the analogue of the Feynman gauge in relativistic gauge theories.

The analogue of the Lorenz gauge in relativistic gauge theories for the Aristotelian case has been studied in  \cite{Anselmi, Blas, pHL}, which we briefly review here. We choose the following nonsingular gauge-fixing functional,
\be
    f[a_\mu] = \dot{a}_0 - (- \p^2 + c_a^2) \, \p_i a_i.
\ee
The Faddeev-Popov action contains a gauge-fixing term,
\be
    S_\text{g.f.} = - \frac{1}{2} \int dt \, d^3\by \, f [a_\mu] \, \Xi^{-1} f[a_\nu],
    \label{eq:RxiSaSgf}
\ee
where $\Xi$ is a dimensionful operator, $[\Xi] = 1$. Consequently, we obtain the following gauge-fixed action,
\be
    S_a + S_\text{g.f.}
    = \frac{1}{2} \int dt \, d^3 \by \, 
        \begin{pmatrix}
            a_0 & a_i
        \end{pmatrix}
        \begin{pmatrix}
            M_{00} & M_{0i} \\
            M_{0j} & M_{ij}
        \end{pmatrix}
        \begin{pmatrix}
            a_0 \\ 
            a_i
        \end{pmatrix}, \label{eq: Matrix}
\ee
where
\begin{subequations}
\begin{align}
    M_{00} & = \frac{\p_0^2}{\Xi} - \p^2, \\
    M_{0i} & = \p_i \p_0 \lr 1 - \frac{-\p^2 + c_a^2}{\Xi} \rr, \\
    M_{ij} & = - \p_0^2 \delta_{ij} + (-\p^2 + c_a^2) \ls \delta_{ij} \p^2 - \lr 1 - \frac{-\p^2+c_a^2}{\Xi} \rr \p_i \p_j \rs.
\end{align}
\end{subequations}
It is convenient to choose 
\be
    \Xi = \xi (- \p^2 + c_a^2),
    \label{eq:RxiXi}
\ee
where $\xi$ is a dimensionless gauge-dependent parameter. Therefore,
\begin{subequations}
\begin{align}
    M_{00} & = \frac{\p_0^2 - \xi (-\p^2 + c_a^2) \p^2}{\xi (- \p^2 + c_a^2)}, \label{eq:M00} \\[2pt]
    M_{0i} & = \lr 1 - \xi^{-1} \rr \p_i \p_0, \\[2pt]
    M_{ij} & = \ls - \p_0^2 + (-\p^2 + c_a^2) \p^2 \rs \delta_{ij} - \lr 1 - \xi^{-1} \rr \p_i \p_j (-\p^2 + c_a^2).
\end{align}
\end{subequations}
The nonlocal behavior of \eqref{eq:M00} can be eliminated by taking the field redefinition \eqref{eq:fda0}. Note that, after this rescaling, $a_0$ and $a_i$ have the same scaling dimension, \emph{i.e.}, $[a_0] = [a_i] = 1/4$. 

Inverting the matrix defined in \eqref{eq: Matrix}, we obtain the gauge propagator in the Lorenz gauge (with the choice of $\Xi$ in \eqref{eq:RxiXi}),
\be
    \Delta^{\mu\nu}_{\text{Lorenz}} = \frac{-i}{K^2 + i \varepsilon} \ls
        \begin{pmatrix}
            (|\bk|^2 + c_a^2) & 0 \\
            0 & - \delta_{ij}
        \end{pmatrix}
            - (1-\xi) \frac{|\bk|^2 + c_a^2}{K^2 + i \varepsilon} 
        \begin{pmatrix}
            \omega^2 & - \omega k_i \\
            - \omega k_j & k_i k_j
        \end{pmatrix}
        \rs
\ee
When $\xi = 1$, the Lorenz gauge propagator reduces to \eqref{eq:Deff}. Moreover, performing the field redefinition \eqref{eq:fda0}, we obtain
\be \label{eq:xiprop}
    \Delta^{\mu\nu}_\text{Lorenz} \rightarrow \frac{-i}{K^2 + i \varepsilon} \ls \eta^{\mu\nu} - (1-\xi) \frac{K^\mu K^\nu}{K^2 + i \varepsilon} \rs. 
\ee
This is in complete analogy with the relativistic Lorenz gauge propagator in \eqref{eq:Rxirel}, except that $k_\mu$ is replaced by $K_\mu = (\omega, \bk \sqrt{|\bk|^2 + c_a^2})$.

In the following, we will always perform the field redefinition \eqref{eq:fda0}, which will affect the Feynman rules for both the gauge propagator and the vertices. We have derived the corresponding gauge propagator in Lorenz gauge in \eqref{eq:xiprop}. In Coulomb gauge, the propagator \eqref{eq:coulombz2} becomes
\be
\label{eq:Coulombpropagatorz=2}
\Delta^{\mu\nu}_\text{Coul} (k) = \begin{pmatrix}
            \frac{i}{|\bK|^2} & 0 \\
            0 & \frac{i}{K^2 + i \epsilon}\left( \delta_{ij} - \frac{K_i K_j}{|\bK|^2}\right)
        \end{pmatrix}.
\ee
Formally, this expression looks very similar to \eqref{eq:reCgp} in the relativistic case.

\subsection{Feynman rules}

In the rest of the section, we calculate the gauge field one-loop correction to $m^2$ and $c^2$ in both Coulomb and Lorenz gauge.
Henceforth, we will set $c_a = c = 0$ and $m=0$ since these regulate IR divergences, whereas we are interested in UV divergences. The three-point vertex is given by
\begin{subequations} \label{eq:3ptz=2}
\begin{align}
\begin{minipage}{3.3cm}
\begin{tikzpicture}
\draw[decorate,decoration=snake] (0,-0.5) -- (0,0.5) ;
\draw[->] (-1,-1.5) -- (-0.5,-1) ;
\draw[->][thick] (-0.5,-1) -- (-0.49,-0.99) ;
\draw (-0.5,-1) -- (0,-0.5) ;
\draw[->] (1,-1.5) -- (0.5,-1) ;
\draw[->][thick] (0.5,-1) -- (0.49,-0.99) ;
\draw (0.5,-1) -- (0,-0.5) ;
\draw  [->][thick]  (20:-0.08)   -- + (250:0.01) ;
\small \node [below] at (-1,-1.5) {$\oln{\phi}\left(\qk_1\right)$};
\node [below] at (1,-1.5) {$\phi\left(\qk_2\right)$};
\small \node [above] at (0,0.5) {$a_{\mu}(\qq)$};
\end{tikzpicture}
\end{minipage}
&\equiv ie V_{\mu} (q,k_1,k_2) = ie \binom{( \omega_2 - \omega_1 ) | \bq |}{\bV (q,k_1,k_2)}, \label{eq:Vmuz=2} \\
\intertext{where}
\bV (q,k_1,k_2) &= ( | \bk_1 |^2 + | \bk_2 |^2 ) ( \bk_2 - \bk_1 ) + \eta_1 \bigl[ ( \bq \cdot \bk_1 ) \bk_2 - ( \bq \cdot \bk_2 ) \bk_1 \bigr].
\end{align}
\end{subequations}
We actually need two copies of this vertex evaluated with specific momenta, $V_{\mu} (q,k,-q-k)$ and $V_{\mu} (-q,q+k,-k)$. One can easily check that these are in fact identical vertices,
\begin{align} \label{eq:V}
    & \quad \bV (q, k, -q, -k) = \bV (-q, q+k, -k) \notag \\
    & = - |\bq|^2 \bq - \Bigl[ (2-\eta_1) |\bq|^2 \bk + (2+\eta_1) (\bq \cdot \bk) \bq \Bigr] - \Bigl[ 2 |\bk|^2 \bq + 4 (\bq \cdot \bk) \bk \Bigr] + \CO (|\bk|^3).
\end{align}
We will not actually need the full four-point vertex, but just the one evaluated on the following specific frequencies and momenta:
\begin{subequations} \label{eq:Vmunuz=2}
\begin{align}
\begin{minipage}{3.3cm}
\begin{tikzpicture}
\draw[decorate,decoration=snake] (0,-0.5) -- (1,0.5) ;
\draw[decorate,decoration=snake] (0,-0.5) -- (-1,0.5) ;
\draw[->] (-1,-1.5) -- (-0.5,-1) ;
\draw[->][thick] (-0.5,-1) -- (-0.49,-0.99) ;
\draw (-0.5,-1) -- (0,-0.5) ;
\draw[->] (1,-1.5) -- (0.5,-1) ;
\draw[->][thick] (0.5,-1) -- (0.49,-0.99) ;
\draw (0.5,-1) -- (0,-0.5) ;
\draw  [->][thick]  (0:0.5)   -- + (260:0.01) ;
\draw  [->][thick]  (0:-0.5)   -- + (360:0.01);
\small \node [below] at (-1,-1.5) {$\oln{\phi}\left(\qk\right)$};
\node [below] at (1,-1.5) {$\phi\left(-\qk\right)$};
\small \node [above] at (1,0.5) {$a_{\nu}(-\qq)$};
\small \node [above] at (-1,0.5) {$a_{\mu}(\qq)$};
\end{tikzpicture}
\end{minipage}
&= 2ie^2 V_{\mu\nu} (q,k) = 2ie^2
\begin{pmatrix}
    | \bq |^2 & 0 \\
    0 & V_{ij} (q,k)
\end{pmatrix}, \\
\intertext{where}
V_{ij} (q,k) &= -q_i q_j - 4k_i k_j - 2 | \bk |^2 \delta_{ij} - \eta_2 ( | \bq |^2 \delta_{ij} - q_i q_j ).
\end{align}
\end{subequations}
Combined with the propagators in \eqref{eq:coulombpropeff} and \eqref{eq:xiprop} and the scalar propagator in \eqref{eq:scalarprop}, these suffice to calculate the one-loop corrections to the scalar propagator in both Coulomb and Lorenz gauges.

\subsection{One-loop correction in Coulomb gauge} \label{sec:Cgz2}

Let us first compute the one-loop correction to the scalar propagator in Coulomb gauge. We will take a Taylor expansion of the Feynman diagrams with respect to the external momentum $\bk$ and only keep up to $\CO(|\bk|^2)$. This is sufficient for us to extract useful information about quantum corrections to $m^2$ and $c^2$ in the scalar sector. For our interest in two-point correlation functions at one-loop, it is always possible to use the conservation law of momentum to write the Feynman diagrams in a way such that only scalar internal legs contain external momenta. Hence, our Taylor expansions will always be taken with respect to the smallness of $|\bk| / \sqrt{m}$ as in \cite{irs}. In the following calculation, however, we will not write out $m$ explicitly. 

In Coulomb gauge, the cog diagram is given by
\begin{equation}
\begin{minipage}{7cm}
\begin{tikzpicture}
\draw  [decorate,decoration=snake] (0,0.77) circle (0.77);
\draw  [->][thick]  (82:1.63)  -- + (20:-0.01);
\draw   (-1.7,0) -- (1.75,0);
\draw [->] [thick]   (-1.21,0) -- (-1.2,0);
\draw [<-]  [thick] (1.3,0) -- (1.31,0);
\node [left] at (-1.2,-0.4) {$\overline{\phi} (\qk)$};
\node [right] at (1.2,-0.4) {$\phi (-\qk)$};
\node [above][ultra thick] at (0.1,1.65) {$\qq$};
\end{tikzpicture}
\end{minipage}
\hspace{-2cm}
= \frac{1}{2} \int \frac{d^4 q}{(2\pi)^4} \, 2i e^2 V_{\mu\nu} (\qq, \qk) \, \Delta^{\mu\nu}_\text{Coul} (q),
\label{eq:cogCoulz2}
\end{equation}
where $\Delta^{\mu\nu}_\text{Coul}$ is given by \eqref{eq:Coulombpropagatorz=2}.
This cog diagram requires the trace of $V_{ij}$,
\begin{equation} \label{eq:trV}
    V_{i}^{i} (q,k) = - (1 + 2 \eta_2 ) | \bq |^2 - 10 | \bk |^2,
\end{equation}
and the combination
\begin{equation*}
    \frac{q_i V_{ij} q_j}{| \bq |^2} = - | \bq |^2 - 2 | \bk |^2 - \frac{4 ( \bq \cdot \bk )^2}{| \bq |^2}.
\end{equation*}
Since these expressions are eventually integrated over $q$, and we are keeping only up to $\CO(|\bk|^2)$ (higher order terms contribute UV-finite integrals), we can always replace $( \bq \cdot \bk )^2$ with $\frac{1}{3} | \bq |^2 | \bk |^2$. We indicate this replacement with the symbol ``$\rightarrow$". Thus,
\begin{equation} \label{eq:qVq}
    \frac{q_i V_{ij} q_j}{| \bq |^2} \rightarrow - | \bq |^2 - \frac{10}{3} | \bk |^2,
\end{equation}
and
\begin{equation}
    V_{ij} (q,k) \biggl( \delta_{ij} - \frac{q_i q_j}{| \bq |^2} \biggr) \rightarrow -2 \eta_2 | \bq |^2 - \frac{20}{3} | \bk |^2.
\end{equation}
The cog diagram \eqref{eq:cogCoulz2} evaluates to
\begin{equation*}
    i \Gamma_{\text{cog}} = e^2 \int \frac{d \nu}{2 \pi} \frac{d^3 \bq}{( 2 \pi )^3} \biggl( \frac{2 \eta_2 | \bq |^2 + \frac{20}{3} | \bk |^2}{\nu^2 - | \bq |^4} - \frac{1}{| \bq |^2} \biggr).
\end{equation*}
We write this as
\begin{equation} \label{eq:cogz=2}
    \Gamma_{\text{cog}} = -2 \eta_2 e^2 \mathcal{I}_{1}^{(2)} - \frac{20}{3} e^2 \mathcal{J}_{1}^{(2)} | \bk |^2 + ie^2 \int \frac{d \nu}{2 \pi} \frac{d \bq}{( 2 \pi )^3} \frac{1}{| \bq |^2},
\end{equation}
where 
\begin{subequations} \label{eq:divints}
\begin{align}
\mathcal{I}_{n}^{(2)} &\equiv i \int \frac{d \qn}{2\pi} \frac{d^3 \bqq}{(2\pi)^3} \frac{|\bqq|^2 | \bq |^{4(n-1)}}{\bigl( \qn^2 - |\bqq|^{4} \bigr)^n}, \\
\mathcal{J}_{n}^{(2)} &\equiv i \int \frac{d \qn}{2\pi} \frac{d^3 \bqq}{(2\pi)^3} \frac{| \bq |^{4(n-1)}}{\bigl( \qn^2 - |\bqq|^{4} \bigr)^n}. \label{eq:defJn2}
\end{align}
\end{subequations}

The sunset diagram is given by
\begin{equation} \label{eq:sunsetz=2}
\begin{minipage}{6cm}
\begin{tikzpicture}
\begin{scope}
\clip (-1,0) rectangle (1,1);
\draw  [decorate,decoration=snake] (0,0) circle (0.75);
\end{scope}
\draw  [->][thick]  (0.0,0.82)  -- (-.01,.83) ;
\draw  [<-][thick]  (250:0)  -- + (0:0.01) ;
\draw  (-1.7,0) -- (1.75,0);
\draw [->]  [thick] (-1.21,0) -- (-1.2,0);
\draw [<-]  [thick] (1.3,0) -- (1.31,0);
\node [left] at (-1.2,-0.4) {$\overline{\phi} (\qk)$};
\node [right] at (1.2,-0.4) {$\phi (-\qk)$};
\node [above] at (0.1,0.9) {$\qq$};
\node [below] at (0.1,-0.1) {$\qp$};
\end{tikzpicture}
\end{minipage}
\hspace{-2cm}
= \int \frac{d^4 q}{(2\pi)^4} \, ie V_{\mu}(q,k,p) \, \Delta^{\mu \nu}_\text{Coul}(q) \, ie V_{\nu} (-q,-p,-k) \, \Delta(p),
\end{equation}
This sunset diagram requires the square magnitude of $\bV$,
\be \label{eq:V2}
    | \bV (q,k,-q-k) |^2 
    = | \bq |^4 \biggl( | \bq |^2 + 8 \, \bq \cdot \bk  + \frac{44 - 8 \eta_1 + 2 \eta_{1}^{2}}{3} | \bk |^2 \biggr) + \CO (|\bk|^3),
\ee
as well as the combination
\begin{align}
    \frac{( \bq \cdot \bV )^2}{| \bq |^2} 
    \rightarrow | \bq |^4 \Bigl[ | \bq |^2 + 8 ( \bq \cdot \bk ) + 12 | \bk |^2 \Bigr] + \CO (|\bk|^3).
\end{align}
Therefore,
\begin{equation} \label{eq:qVqV}
    V_i \biggl( \delta_{ij} - \frac{q_i q_j}{| \bq |^2} \biggr) V_j = | \bV |^2 - \frac{( \bq \cdot \bV )^2}{| \bq |^2} \rightarrow \frac{8 - 8 \eta_1 + 2 \eta_{1}^{2}}{3} | \bq |^4 | \bk |^2.
\end{equation}
The sunset diagram \eqref{eq:sunsetz=2}
evaluates to
\begin{align}
    i \Gamma_{\text{sunset}} &= e^2 \int  \frac{d \nu}{2 \pi} \frac{d^3 \bq}{(2 \pi )^3} \frac{1}{\nu^2 - | \bq + \bk |^4} \biggl( \frac{\nu^2}{| \bq |^2} + \frac{8 - 8 \eta_1 + 2 \eta_{1}^{2}}{3} \frac{| \bq |^4 | \bk |^2}{\nu^2 - | \bq |^4} \biggr),
\end{align}
where we have set the external frequency $\omega$ to zero since the correction to $\omega^2$ is finite. We must expand the propagator when it multiplies $\nu^2 / | \bq |^2$,
\begin{align} \label{eq:scalarpropexp}
    \frac{1}{\nu^2 - | \bq + \bk |^4} 
    \rightarrow \frac{1}{\nu^2 - | \bq |^4} \biggl[ 1 + \frac{4| \bq |^2 ( \bq \cdot \bk ) + \frac{10}{3} | \bq |^2 | \bk |^2}{\nu^2 - | \bq |^4} + \frac{\frac{16}{3} | \bq |^6 | \bk |^2}{\bigl( \nu^2 - | \bq |^4 \bigr)^2} \biggr] + \CO(|\bk|^3).
\end{align}
We will drop the $\CO(|\bk|)$ term because it gives an odd integrand and thus vanishes.
We find
\begin{align}
    \Gamma_{\text{sunset}} &= - e^2 \mathcal{I}_{1}^{(2)} - e^2 \biggl( \frac{10}{3} \mathcal{J}_{1}^{(2)} + \frac{34 - 8 \eta_1 + 2 \eta_{1}^{2}}{3} \mathcal{J}_{2}^{(2)} + \frac{16}{3} \mathcal{J}_{3}^{(2)} \biggr) | \bk |^2 \notag \\
    &\quad -ie^2 \int \frac{d \nu}{2 \pi} \frac{d^3 \bq}{( 2 \pi )^3} \frac{1}{| \bq |^2}. 
    \label{eq:Gammasun}
\end{align}

Combining \eqref{eq:Gammasun} with \eqref{eq:cogz=2} gives
\begin{align} \label{eq:Gammaz=2}
    \Gamma &= \Gamma_{\text{cog}} + \Gamma_{\text{sunset}} \notag \\
    &= - (1+ 2 \eta_2 ) e^2 \mathcal{I}_{1}^{(2)} - \frac{e^2}{3} \Bigl[ 30 \mathcal{J}_{1}^{(2)} + \lr 34 - 8 \eta_1 + 2 \eta_{1}^{2} \rr \mathcal{J}_{2}^{(2)} + 16 \mathcal{J}_{3}^{(2)} \Bigr] | \bk |^2.
\end{align}
Taking into account the correction from $\Gamma$ to the scalar propagator \eqref{eq:scalarpropz2}, the exact propagator is 
\begin{equation}
    \frac{i}{\omega^2 - |\bk|^4 - c^2 |\bk|^2 - m^2 + \Gamma}.
\end{equation}
Therefore,
\begin{subequations}
\begin{align}
    \delta m^2 & = (1+2\eta_2) e^2 \CI^{(2)}_1, \\[2pt]
    \delta c^2 & = \frac{e^2}{3} \Bigl[ 30 \mathcal{J}_{1}^{(2)} + \lr 34 - 8 \eta_1 + 2 \eta_{1}^{2} \rr \mathcal{J}_{2}^{(2)} + 16 \mathcal{J}_{3}^{(2)} \Bigr]. \label{eq:deltac2Cg}
\end{align}
\end{subequations}
Note that for $\eta_2 = -1/2$, the $e^2$ order correction to $m^2$ vanishes. The linear divergence in $\delta c^2$ is conceptually trickier to deal with and will be discussed in detail in Section \ref{sec:pld}. 
 
\subsection{One-loop correction in Lorenz gauge} \label{sec:Rxiz2}

In Lorenz gauge, the cog diagram in \eqref{eq:cogz=2} evaluates to
\begin{align*}
    i \Gamma_{\text{cog}} &= e^2 \int \frac{d \nu}{2 \pi} \frac{d^3 \bq}{(2 \pi )^3} \frac{1}{\nu^2 - | \bq |^4} \biggl[ V_{\mu}^{\mu} - (1- \xi ) \frac{| \bq |^2 ( \nu^2 + q_i V_{ij} q_j )}{\nu^2 - | \bq |^4} \biggr] \notag \\
    &= e^2 \int \frac{d \nu}{2 \pi} \frac{d^3 \bq}{(2 \pi )^3} \frac{1}{\nu^2 - | \bq |^4} \biggl[ (1 + 2 \eta_2 + \xi ) | \bq |^2 + 10 | \bk |^2 + \frac{10}{3} (1 - \xi ) \frac{| \bq |^4 | \bk |^2}{\nu^2 - | \bq |^4} \biggr],
\end{align*}
where use was made of \eqref{eq:trV} and \eqref{eq:qVq}. 
Hence,
\begin{equation} \label{eq:cogxiz=2}
    \Gamma_{\text{cog}} = - \bigl(1 + 2 \eta_2 + \xi \bigr) e^2 \mathcal{I}_{1}^{(2)} - \frac{10}{3} e^2 \bigl[ 3 \mathcal{J}_{1}^{(2)} + (1 - \xi ) \mathcal{J}_{2}^{(2)} \bigr] | \bk |^2.
\end{equation}

The sunset diagram in \eqref{eq:sunsetz=2} evaluates to
\begin{align*}
    i \Gamma_{\text{sunset}} &= -e^2 \int \frac{d \nu}{2 \pi} \frac{d^3 \bq}{(2 \pi )^3}  \frac{1}{\nu^2 - | \bq + \bk |^4} \frac{1}{\nu^2 - | \bq |^4}  \times\biggl[ V_{\mu} V^{\mu} - (1- \xi ) \frac{( Q^{\mu} V_{\mu} )^2}{\nu^2 - | \bq |^4} \biggr].
\end{align*}
Using \eqref{eq:V2} and \eqref{eq:qVqV}, we find
\begin{subequations}
\begin{align}
    \frac{V_{\mu} V^{\mu}}{\nu^2 - | \bq |^4}
    &\rightarrow | \bq |^2 \biggl( 1 - \frac{8 | \bq |^2 ( \bq \cdot \bk )}{\nu^2 - | \bq |^4} - \frac{44 -8 \eta_1 + 2 \eta_{1}^{2}}{3} \frac{| \bq |^2 | \bk |^2}{\nu^2 - | \bq |^4} \biggr), \label{eq:VV} \\[2pt]
    \frac{( Q^{\mu} V_{\mu} )^2}{( \nu^2 - | \bq |^4 )^2} 
    &\rightarrow | \bq |^2 \biggl[ 1 - \frac{8 | \bq |^2 ( \bq \cdot \bk )}{\nu^2 - | \bq |^4} - \frac{\frac{20}{3} | \bq |^2 | \bk |^2}{\nu^2 - | \bq |^4} + \frac{\frac{16}{3} | \bq |^6 | \bk |^2}{( \nu^2 - | \bq |^4 )^2} \biggr]. \label{eq:QV}
\end{align}
\end{subequations}
Subtracting \eqref{eq:QV} from \eqref{eq:VV}, multiplying by \eqref{eq:scalarpropexp}, and plugging back into the expression for $\Gamma_{\text{sunset}}$ gives
\begin{align}
    \Gamma_{\text{sunset}} 
    = \xi e^2 \mathcal{I}_{1}^{(2)} & - \frac{24 - 8 \eta_1 + 2 \eta_{1}^{2} + 10 \xi}{3} e^2 \mathcal{J}_{2}^{(2)} | \bk |^2 - \frac{16}{3} e^2 \mathcal{J}_{3}^{(2)} | \bk |^2.
\end{align}
Adding this to \eqref{eq:cogxiz=2} gives
\begin{align} \label{eq:Gammaxiz=2}
    \Gamma &= \Gamma_{\text{shoelace}} + \Gamma_{\text{sunset}} \notag \\
    &= - (1+ 2 \eta_2 ) e^2 \mathcal{I}_{1}^{(2)} - \frac{e^2}{3} \Bigl[ 30 \mathcal{J}_{1}^{(2)} + \lr 34 - 8 \eta_1 + 2 \eta_{1}^{2} \rr \mathcal{J}_{2}^{(2)} + 16 \mathcal{J}_{3}^{(2)} \Bigr] | \bk |^2.
\end{align}
This is independent of $\xi$ and is identical to the Coulomb gauge result \eqref{eq:Gammaz=2}.

\subsection{Power law divergences in the Aristotelian spacetime}
\label{sec:pld}

In Section \ref{sec:Cgz2} and \ref{sec:Rxiz2} we find that the leading quantum correction to $c^2$ is given by \eqref{eq:deltac2Cg} in both Coulomb and Lorenz gauges, 
\be
    \delta c^2 = \frac{2 e^2}{3} \lc 15 \, \mathcal{J}^{(2)}_{1} + \ls 13 + \lr \eta_1 - 2 \rr^2 \rs \mathcal{J}^{(2)}_{2} + 8 \, \mathcal{J}^{(2)}_{3} \rc,
    \label{eq:deltac2pld}
\ee
where
\begin{align}
    \mathcal{J}^{(2)}_n = i \int \frac{d\nu}{2\pi} \frac{d^3\bq}{(2\pi)^3} \frac{|\bq|^{4(n-1)}}{(\nu^2 - |\bq|^4)^n},
\end{align}
as defined in \eqref{eq:defJn2}.
All coefficients of $\mathcal{J}_n^{(2)}$, $n = 1, 2, 3$ are individually gauge independent in \eqref{eq:deltac2pld}. After Wick rotating to imaginary time, we obtain
\be
    \mathcal{J}^{(2)}_n = (-1)^{n+1} \int \frac{d\nu}{2\pi} \frac{d^3\bq}{(2\pi)^3} \frac{|\bq|^{4(n-1)}}{(\nu^2+|\bq|^4)^n},
    \label{eq:mathfrakJn}
\ee
which is linearly divergent in momentum. To define such power law divergences, first we need to define how the integrals are regulated in the UV. In a relativistic theory, space and time components are related to each other by Lorentz boosts, and, consequently, the UV sharp cutoff at a given energy scale $\Lambda$ is uniquely determined to be a round sphere of radius $\Lambda$, centered at the origin in frequency-momentum space. In  Aristotelian spacetime, however, there is no preferred UV regulation scheme. In general, a hypersurface in the $\nu$-$\bq$ space that respects the reflection symmetry $\nu \rightarrow -\nu$ and rotational symmetry in $\bq$ is a valid choice. Note, however, that in both the Aristotelian case and the relativistic case, the power law divergences functionally depend on the cutoff. The only subtlety in our case is a technical one: the functional dependence is on the shape of the cutoff surface, rather than just one scale. 

Are there any universal relations among $\mathcal{J}^{(2)}_n$'s that are insensitive to the detailed shape of the UV cutoff surface? Interestingly, there is no such relation and $\mathcal{J}^{(2)}_n$'s appear to describe independent divergences, even though by power counting they are all linearly divergent at the $z=2$ Gaussian fixed point. 

With the explicit expressions of the integrals \eqref{eq:mathfrakJn} in hand, it is quite simple to see their sensitivity to the shape of the UV cutoff. Let us take a simple example and require the UV cutoff hypersurface in the $\nu$-$\bq$ space to be a cylinder, such that the integral \eqref{eq:mathfrakJn} is defined over the following domain:
\be
    \Big \{ (\nu, \bq) : \, -\Omega < \nu < \Omega, \, 0 < |\bq| < \Lambda \Big \},
    \label{eq:CIdomain}
\ee
where $\Omega, \Lambda > 0$. Around the $z=2$ Gaussian fixed point, we have $[\Omega] = 1$ and $[\Lambda] = 1/2$. The integral in \eqref{eq:mathfrakJn} over the domain \eqref{eq:CIdomain} can be performed analytically, giving
$$
    \mathcal{J}^{(2)}_n \! = \! \Omega^\frac{1}{2} \frac{(-1)^{n+1}}{2 \pi^3} \lc \frac{\Gamma (\tfrac{1}{4}) \, \Gamma (n-\tfrac{1}{4})}{2\Gamma(n)} + \frac{\Omega^\frac{1}{2}}{\Lambda} \! \ls {}_2F_1 \lr\tfrac{1}{2}, n; \tfrac{3}{2}; - \tfrac{\Omega^2}{\Lambda^4}\rr - 2{}_2F_1 \lr\tfrac{1}{4}, n; \tfrac{5}{4}; - \tfrac{\Omega^2}{\Lambda^4}\rr \rs \rc,\!\!
$$
where ${}_2F_{1} (a,b;c;d)$ is the hypergeometric function. Instead of working directly with hypergeometric functions, let us focus on two asymptotic limits in which $\mathcal{J}_n^{(2)}$ simplifies
\be
    \mathcal{J}^{(2)}_n = \frac{(-1)^{n+1}}{4\pi^3} \times
        \begin{cases}
            \frac{\Gamma\lr n-\tfrac{1}{4}\rr \, \Gamma\lr\tfrac{1}{4}\rr}{\Gamma(n)} \sqrt{\Omega} - \frac{2 \Omega}{\Lambda} + \cdots, & \Omega \ll \Lambda^2, \\[3pt]
            \frac{\sqrt{\pi} \Gamma \lr n-\tfrac{1}{2}\rr}{\Gamma(n)} \Lambda - \frac{2}{(4n-1) (2n-1)} \lr \frac{\Lambda^2}{\Omega} \rr^{\!2n} \frac{\Omega}{\Lambda} + \cdots, & \Omega \gg \Lambda^2.
        \end{cases}
\ee
There are different limits that we take to push the UV cutoffs to infinity, which define very distinct notions of power law divergences. Here are several examples:
\begin{itemize}

\item

First take $\Omega \rightarrow \infty$ and then take $\Lambda \rightarrow \infty$. Taking the limit $\Omega \rightarrow \infty$, we obtain 
\be \label{eq:reg1}
    \mathcal{J}^{(2)}_n = \frac{(-1)^{n+1} \sqrt{\pi} \, \Gamma \lr n-\tfrac{1}{2}\rr}{4\pi^3\Gamma(n)} \Lambda,
\ee
which is linearly divergent in $\Lambda$.

\item First take $\Lambda \rightarrow \infty$ and then take $\Omega \rightarrow \infty$. Taking the limit $\Lambda \rightarrow \infty$, we obtain
\be 
    \mathcal{J}^{(2)}_n = \frac{(-1)^{n+1} \Gamma\lr n-\tfrac{1}{4} \rr \Gamma\lr \tfrac{1}{4} \rr}{4\pi^3\Gamma\lr n\rr} \sqrt{\Omega},
    \label{eq:sqrtOmega}
\ee
which is linearly divergent in $\sqrt{\Omega}$. 

\item

Simultaneously take $\Omega$ and $\Lambda$ to infinity along a curve $\omega = \alpha k^2$, with $\alpha > 0$ fixed. In this case, $\Lambda$ will be related to $\Omega$ by $\Omega = \alpha \Lambda^2$. In this limit, we obtain
\begin{align*}
    & \mathcal{J}^{(2)}_n \! = \! \Lambda \frac{(-1)^{n+1}}{2 \pi^3} \Bigl \{ \frac{\sqrt{\alpha} \, \Gamma (\tfrac{1}{4}) \, \Gamma (n-\tfrac{1}{4})}{2\Gamma(n)} + \alpha \bigl[ {}_2F_1 \lr\tfrac{1}{2}, n; \tfrac{3}{2}; - \alpha^2 \rr - 2{}_2F_1 \lr\tfrac{1}{4}, n; \tfrac{5}{4}; - \alpha^2 \rr \bigr] \Bigr \},\!\!
\end{align*}
which is linearly divergent in $\Lambda$ and the coefficient of this linear divergence is $\alpha$-dependent.

\end{itemize}
It is clear that there are \emph{a priori} no relations among $\mathcal{J}_n^{(2)}$'s that are independent of how one chooses the UV regulation. For example, using the UV regulation described in \eqref{eq:reg1}, one finds $\mathcal{J}_1^{(2)} / \mathcal{J}_2^{(2)} = -2$, whereas \eqref{eq:sqrtOmega} gives $\mathcal{J}_1^{(2)} / \mathcal{J}_2^{(2)} = -4/3$. As a result, $\mathcal{J}_n^{(2)}$ should be treated as unrelated power law divergences and need to vanish individually for $\delta c^2$ to be hierarchically smaller than $e^2 M^{2/3}$. This is not possible in \eqref{eq:deltac2pld}.

For logarithmic divergences, however, the relevant integral evaluates to a universal expression $C \log (1/\epsilon)$, where $1/\epsilon$ is the characteristic size of the integration domain and the coefficient $C$ is independent of how one chooses the UV regulation. This implies that the beta functions for theories in the Aristotelian spacetime around a Gaussian fixed point are still well defined.
Relevant discussions about log divergences can be found in Appendix \ref{app:logdiv}.

\section{Aristotelian Scalar QED with a \texorpdfstring{$z=3$}{z=3} Scaling} \label{sec:QED_z3}

In this section, we consider the covariantization of the complex scalar $\phi$ whose dynamics is defined by the action \eqref{eq:scalaraction}, exhibiting a $z=3$ scaling at high energies. Let us start by setting the marginal coupling $\lambda_3$ to zero and consider the following action,
\begin{align*}
S_{\phi, e=0} & = \int dt \, d^3\textbf{y} \Big [ \p_0 \oln{\phi} \, \p_0 \phi - \zeta_3^2 \p_i \p^2 \oln{\phi} \, \p_i \p^2 \phi - \zeta_2^2 \p^2 \oln{\phi} \, \p^2 \oln{\phi} - c^2 \p_i \oln{\phi} \, \p_i \phi - m^2 \oln{\phi} \phi - \lambda_0 (\oln{\phi} \phi)^2 \Big ].
\end{align*}
We have seen in the previous section that there are three different ways of covariantizing the $\zeta_2^2$ term. There are, however, fifteen ways of covariantizing the $\zeta_3^2$ operator. To classify such operators, we first list all possible contractions among the indices in the operators
\begin{align} \label{eq:Thetadef}
   \Theta_{ijk} &\equiv D_i D_j D_k \phi, &%
   \oln{\Theta}_{ijk} &\equiv \overline{D_i D_j D_k \phi}.
\end{align}
There are in total eleven different ways of contracting the indices in $\Theta_{ijk}$ and $\oln{\Theta}_{ijk}$, seven of which are real,
\begin{subequations} \label{eq:01to12}
\begin{align} 
    \mathcal{O}_1 &\!=\! \oln{\Theta}_{ijk} \Theta_{ijk}, &
    \!\! \mathcal{O}_2 &\!=\! \oln{\Theta}_{ijk} \Theta_{ikj}, \\
    \!\! \mathcal{O}_3 &\!=\! \oln{\Theta}_{ijk} \Theta_{jik}, &
    \mathcal{O}_4 &\!=\! \oln{\Theta}_{ijk} \Theta_{kji}, \\
    \!\! \mathcal{O}_5 &\!=\! \oln{\Theta}_{iik} \Theta_{jjk}, &%
    \!\! \mathcal{O}_6 &\!=\! \oln{\Theta}_{kii} \Theta_{kjj}, \\
    \mathcal{O}_{7} &\!=\! \oln{\Theta}_{iki} \Theta_{jkj}.
\end{align}
\end{subequations}
There are eight more real operators that can be constructed from the remaining four contractions, namely,
\begin{subequations}
\begin{align} 
    \mathcal{O}_8 & = \frac{1}{2} \lr \oln{\Theta}_{ijk} \Theta_{jki} + c.c. \rr=\text{Re} \left( \oln{\Theta}_{ijk} \Theta_{jki} \right), \\
    \mathcal{O}_{9} &= \frac{1}{2} \lr \oln{\Theta}_{iik} \Theta_{jkj} + c.c. \rr = \text{Re} \left( \oln{\Theta}_{iik} \Theta_{jkj} \right), \\%
    \mathcal{O}_{10} &= \frac{1}{2} \lr \oln{\Theta}_{iik} \Theta_{kjj} + c.c. \rr = \text{Re} \left(\oln{\Theta}_{iik} \Theta_{kjj} \right), \\
    \mathcal{O}_{11} &= \frac{1}{2} \lr \oln{\Theta}_{iki} \Theta_{kjj} + c.c. \rr = \text{Re} \left( \oln{\Theta}_{iki} \Theta_{kjj} \right), \\
    \mathcal{O}_{12} & = \frac{i}{2} \lr \oln{\Theta}_{ijk} \Theta_{jki} - c.c. \rr = - \text{Im} \left( \oln{\Theta}_{ijk} \Theta_{jki} \right), \\
    \mathcal{O}_{13} &= \frac{i}{2} \lr \oln{\Theta}_{iik} \Theta_{jkj} - c.c. \rr = - \text{Im} \left( \oln{\Theta}_{iik} \Theta_{jkj} \right), \\
    \mathcal{O}_{14} &= \frac{i}{2} \lr \oln{\Theta}_{iik} \Theta_{kjj} - c.c. \rr = - \text{Im} \left(\oln{\Theta}_{iik} \Theta_{kjj} \right), \\%
    \mathcal{O}_{15} &= \frac{i}{2} \lr \oln{\Theta}_{iki} \Theta_{kjj} - c.c. \rr = - \text{Im} \left(\oln{\Theta}_{iki} \Theta_{kjj} \right).
\end{align}
\label{eq:12to15}
\end{subequations}
Note that $\oln{\Theta}_{ijk} \Theta_{jki}$ and $\oln{\Theta}_{ijk} \Theta_{kij}$ are complex conjugates of each other. 
If one wishes, one can also rewrite the above operators $\CO_i$ in a different basis in analogy with $\eqref{eq:diffbasis}$ for the $z=2$ case,
\begin{align*}
    & \oln{D_i D_j D_k \phi} \, D_i D_j D_k \phi, \\
    & i F_{ij} \lr \oln{D_i D_k \phi} \, D_j D_k \phi - c.c. \rr, \\
    & i \p_k F_{jk} \lr \oln{D_i \phi} \, D_j D_i \phi - c.c. \rr, \\
    &
    \cdots
\end{align*}
This basis is easier to work with in general. However, for the $z=3$ case, the calculation is already quite involved and we mostly evaluate the Feynman diagrams in \emph{Mathematica}, for which the basis given in equations \eqref{eq:01to12} and \eqref{eq:12to15} is more systematic for keeping track of all possible contributions (without duplicating or missing any terms).

We are interested in the divergent quantum corrections to $c^2$ and $m^2$ from the interactions between $\phi$ and $a_\mu$. Due to the presence of polynomial shift symmetries around the $z=3$ Gaussian fixed point, we can open up a naturally large hierarchy among $\zeta_3^2 \sim 1$, $\zeta_2^2 \sim \varepsilon_2$ and $\lambda_0 \sim \varepsilon_0$ (with the naturalness scale $M$ set to 1 and $\varepsilon_0 \ll \varepsilon_2 \ll 1$). In this case, the leading contributions to $c^2$ and $m^2$ come from the $\zeta_3^2$ term, as we have argued in Section \ref{sec:review}. Hence, it is sufficient for us to focus on the following terms in the covariantized action:
\be
    S_{\phi} = \int dt \, d^3\by \Big (\oln{D_0 \phi} \, D_0 \phi - \sum_{i=1}^{15} \eta_i \CO_{i} - \cdots \Big ),
\ee
where
\be
    \sum_{i=1}^{11} \eta_i = \zeta_3^2 = 1
    \label{eq:sumeta1}
\ee
and ``$\cdots$" includes all the other relevant terms.
We have normalized the spacetime coordinates and the field such that $\zeta_3^2$ is $1$.

We also assume that the gauge sector is around a $z=3$ Gaussian fixed point. The gauge action is 
\begin{align}
S_a = \int dt \, d^3 \textbf{y} \Bigl[ \frac{1}{2}E_i E_i - \frac{1}{4} \lr \zeta_{3,a}^2 \p^2 F_{ij} \p^2 F_{ij} + \zeta_{2,a}^2 \p_k F_{ij} \p_k F_{ij} + c_a^2 F_{ij} F_{ij} \rr \Bigr].
\end{align}
We take the field redefinition in analogy with \eqref{eq:fda0},
\be
    a_0 \rightarrow \sqrt{\p^4 - \zeta_{2, a}^2 \p^2 + c_a^2} \, a_0.
\ee
We also define
\be
    K_i \equiv k_i \sqrt{|\bk|^4 + \zeta_{2,a}^2 |\bk|^2 + c_a^2}.
\ee
Then, in Coulomb gauge with $\p_i a_i = 0$, the gauge propagator is
\be
    \Delta^{\mu\nu}_\text{Coul} (k) = \begin{pmatrix}
            \frac{i}{|\bK|^2} & 0 \\
            0 & \frac{i}{K^2 + i \epsilon}\left( \delta_{ij} - \frac{K_i K_j}{|\bK|^2}\right)
        \end{pmatrix}
\ee
In Lorenz gauge, the gauge propagator is
\be
    \Delta^{\mu\nu}_\text{Lorenz} (k) = \frac{-i}{K^2 + i \varepsilon} \ls \eta^{\mu\nu} - (1-\xi) \frac{K^\mu K^\nu}{K^2 + i \varepsilon} \rs. 
\ee
Again, since we are interested in the UV behavior, we will simply omit the IR regulators $\zeta_{2,a}^2$ and $c_a^2$, and the integrals are understood to be regulated in the IR.

The leading corrections to $c^2$ and $m^2$ are 
\begin{align} 
\delta m^2 & =  e^2 \Bigl [ 1 
+ 2 \lr \eta^{}_1 + \eta^{}_3 + \eta^{}_5 \rr \Bigr ] \mathcal{I}^{(3)}_1,
\label{eq:dm2z3}
\end{align}
with
\be
    \mathcal{I}^{(3)}_1 = i \int \frac{d\nu}{2\pi} \frac{d^3 \bq}{(2\pi)^3} \frac{|\bq|^4}{\nu^2 - |\bq|^6},
\ee
which is quartically divergent, and
\begin{align} 
\delta c^2 & =  12 e^2 \mathcal{J}^{(3)}_3 + \frac{e^2}{6} \ls 114 + (2 + 2\eta^{}_6 + \eta^{}_{10} + \eta^{}_{11})^2 + (\eta^{}_{14} + \eta^{}_{15})^2 \rs \mathcal{J}^{(3)}_2 \notag \\
    & \quad + \frac{e^2}{3} \Big [ 35 + 2 \lr 8 \eta^{}_1 + 5 \eta^{}_2 + 5 \eta^{}_3 - 3 \eta^{}_4 + 7 \eta^{}_5 - 3 \eta^{}_{6} - 5 \eta^{}_7 - 2 \eta^{}_{10} - 5 \eta^{}_{11} \rr \Bigr ] \mathcal{J}^{(3)}_1,
    \label{eq:dc2z3}
\end{align}
with the integrals $\mathcal{J}^{(3)}_n$, $n = 1, 2, 3$ defined to be
\be
    \mathcal{J}^{(3)}_n = i  \int  \frac{d \qn}{2\pi} \frac{d^3 \bqq}{(2\pi)^3} \frac{|\bqq|^{6 n - 4}}{\bigl(\qn^2 - |\bqq|^6\bigr)^{n}},
\ee
which are quadratically divergent. The Coulomb gauge and Lorenz gauge give the same results \eqref{eq:dm2z3} and \eqref{eq:dc2z3}. It is also pleasing to observe that all coefficients in front of $\CJ_n^{(3)}$, $n = 1, 2, 3$ are gauge independent.  

Requiring $\delta m^2 = 0$ in \eqref{eq:dm2z3} gives
\be
   1 + 2 \lr \eta_1 + \eta^{}_3 + \eta^{}_5 \rr = 0.
   \label{eq:condm2z3}
\ee
The relation \eqref{eq:condm2z3} will necessarily require that $\eta_1$, $\eta_3$ or $\eta_5$ be negative, which may result in an unbounded Hamiltonian (both from below and above) and the theory may exhibit vacuum decay. However, if the theory is sufficiently weakly coupled, then it is still perturbatively stable around $\corr{a_\mu} = 0$ and $\corr{\phi} = 0$. This is precisely the case here: both the gauge coupling $e$ and the scalar self-coupling $\lambda_3$ are very small, suppressed due to polynomial shift symmetries. 

There are no universal relations among $\mathcal{J}_n^{(3)}$'s, and thus requiring $\delta c^2 = 0$ in \eqref{eq:dc2z3} will force all coefficients in front of $\mathcal{J}^{(3)}_n$ to be zero. However, this is impossible, since both the coefficients in front of $\mathcal{J}^{(3)}_2$ and $\mathcal{J}^{(3)}_3$ are positive definite. This is the reason why we can only set the quartic divergence in $m^2$ to zero but not push it further to set the quadratic divergence to $c^2$ to zero, as discussed in Section \ref{sec:review}.

\section{Conclusions}

This paper focuses on a series of toy models (scalar QEDs) in which a single massive scalar is coupled to a $U(1)$ gauge boson in ($3+1$)-dimensional Aristotelian spacetime. After reviewing the relativistic case (with $z=1$ scaling), we study the scalar QEDs that exhibit higher dynamical critical exponents with $z=2$ and $z=3$. 

Around a $z=2$ Gaussian fixed point, we consider a superrenormalizable theory that is simple enough to work with, but already exhibits intriguing novelties. In contrast to relativistic theories, the existence of power law divergences does not imply strong UV sensitivity; instead, the sizes of quantum corrections can be suppressed by invoking polynomial shift symmetries on the scalar field. The absence of log divergences allows us to freely choose the marginal parameters in the quadratic terms without violating the principle of technical naturalness. This opens up room for further suppressing power law divergences in the theory. 

We proceed with a systematic investigation of gauge fixing in Aristotelian $U(1)$ gauge theories. We develop the analogues of the Coulomb gauge and Lorenz gauge. Working in both gauges gives a strong check of the results obtained in this paper.

In relativistic theories, power law divergences of the same degree are proportional to each other and their proportionality factors are independent of the choice of UV regularization. In Aristotelian field theories, however, power law divergences develop a more refined structure. Different loop integrals of the same positive superficial degree of divergence are usually not universally related to each other. This is not as surprising as it might sound: since there is no boost symmetry in the Aristotelian case that relates the UV regulators of frequency and momentum (except that the scaling dimensions of the frequency and the momentum are fixed with respect to the given dynamical critical exponent $z$), divergences of a given degree form a multiparameter family, depending on the UV energy scale, how the regulators of frequencies and momenta are related, and the detailed expressions of the loop integrals. Nevertheless, log divergences remain insensitive to how one chooses the UV regulator, and thus the universality of beta functions is preserved. 

For power law divergences of a given degree to vanish, all divergences of different types need to vanish individually. It is reassuring that all coefficients of these power law divergences are indeed gauge independent, as we have checked explicitly.

\smallskip

Finally, in the $z=3$ scalar QED, we compute the one-loop quantum corrections to the scalar propagator. While the leading $e^2$ correction to the scalar mass squared $m^2$ can naturally be set to zero, there is not enough room for further suppressing the speed term, even in the presence of many free parameters. This is a direct consequence of the refined structure of power law divergences. 

\smallskip

The study of the $z=3$ scalar QED has direct phenomenological consequences to the Higgs mass hierarchy problem, if applied to the mechanism proposed in \cite{nrn}. Canceling the leading power law divergences in $m^2$ provides us with an opportunity to improve the naturalness of the model: at least in this toy model, we can maintain a hierarchy of 7 orders of magnitude between $m$ and the naturalness scale $M$ while keeping the Higgs quartic self-coupling, the Yukawa couplings, and the gauge coupling of the same order as in the Standard Model. Our results are obtained for a simple model with $U(1)$ gauge symmetry but can be extended to the Standard Model gauge group $SU(3)_C \times SU(2)_L \times U(1)_Y$, which may weaken the hierarchy but should still allow a sizable ratio $M / m$. Moreover, in this toy model, we have taken the gauge couplings to be of order 0.1, which is realistic in comparison to their Standard Model values around the electroweak scale. However, the hierarchy will be further reduced if we enhance the gauge coupling toward $\sim 0.65$, the value relevant for the $W$ and $Z$ bosons. Furthermore, the method proposed in this paper has its own intrinsic limitation, due to the fact that one cannot further suppress the speed term of the scalar. New ideas will be required to push the hierarchy even further while maintaining naturalness. 
 
We only focused on the unbroken phase of the scalar QEDs. One obvious future direction of study is to extend this to the broken phase and explore nonrelativistic quantum behavior in the context of spontaneous symmetry breaking with gauge symmetries. Furthermore, generalizations to non-Abelian Yang-Mills theories would be a final goal for us to determine whether our mechanism is useful for addressing the Higgs mass hierarchy problem in the Standard Model. This study is not only relevant to Higgs physics as in our original motivation but also should shed some light on the effective field theory of inflation \cite{efti, eftiw}, where a single scalar is coupled to gravity.

\acknowledgments

We thank P.~Ho\v{r}ava, B.~Kol, C.~Melby-Thompson, C.~Mogni, N.~Obers and S.~Sarkar for useful discussions at various stages during this project.
The work of L.B. was supported in part by the Danish National Research Foundation (DNRF91).
The work of K.T.G. was supported in part by ERC Advanced Grant No. 291092 ``Exploring the Quantum Universe" and the Independent Research Fund Denmark project ``Towards a deeper understanding of black holes with nonrelativistic holography."  K.T.G. is grateful for the hospitality of the Racah Institute, Hebrew University of Jerusalem and for funding granted by the Faculty of Science of University of Copenhagen to stimulate Danish-Israeli scientific collaboration.
Z.Y. is grateful for the hospitality of the Niels Bohr Institute, where most of this work was completed, and support from the Danish National Research Foundation project
``New horizons in particle and condensed matter physics from black holes."
The work of Z.Y. was supported in part by NSF Grant No. PHY-1521446, by Berkeley Center for Theoretical Physics and by the Perimeter Institute for Theoretical Physics. Z.Y. thanks the BCTP Brantley-Tuttle fellowship for support while this work was completed. This research was supported in part by Perimeter Institute for Theoretical Physics. Research at Perimeter Institute is supported by the Government of Canada through the Department of Innovation, Science and Economic Development and by the Province of Ontario through the Ministry of Research, Innovation and Science.

\appendix

\section{Universality of Logarithmic Divergences}
\label{app:logdiv}

We illustrate the universality of log divergences with the example of a single real scalar $\phi$ around a $z=2$ Gaussian fixed point \cite{gg,Lubensky,cmu} in a ($2+1$)-dimensional Aristotelian spacetime,
$$
    S = \frac{1}{2} \! \int \! dt \, d^2 \bx \ls \dot{\phi}^2 - \lr \p^2 \phi \rr^2 - c^2 \p_i \phi \p_i \phi - \frac{\lambda}{4} (\p_i \phi \p_i \phi)^2 \rs\!.
$$
This theory enjoys the reflection symmetry $\phi \rightarrow - \phi$ and the constant shift symmetry. At the $z = 2$ Gaussian fixed point, $\lambda$ is classically marginal. The speed term with a $c^2$ coupling is the only relevant term. We are interested in studying how integrating out higher energy modes affects the running of the coupling $\lambda$ in the low-energy effective field theory. 

The Feynman rules of this theory are straightforward to derive. The propagator is
\be \label{eq:prop}
\begin{minipage}{2.6cm}
\begin{tikzpicture}
	\draw [->] (0,0) -- (1,0);
	\draw [->] [thick] (1,0) -- (1.01,0);
	\draw (1,0) -- (2,0);
	\node [white] at (1,-0.3) {$k$};
	\node at (1,0.3) {$k = (\omega, \bk)$};
\end{tikzpicture}
\end{minipage} 
\qquad
\Delta (k) = \frac{i}{\omega^2 - | \bk |^4 - c^2 | \bk |^2}.
\ee
There is also a four-point vertex,
$$
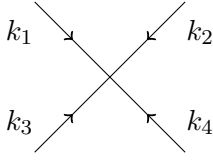
\begin{minipage}{3.7cm}
\begin{tikzpicture}[ scale=0.5]
	\draw [->] (0,0) -- (1,1);
	\draw [->] [thick] (1,1) -- (1.01,1.01);
	\draw (1,1) -- (3,3);
	\draw [->] (4,4) -- (3,3);
	\draw [->] [thick] (3,3) -- (2.99,2.99);
	\draw [->] (4,0) -- (3,1);
         \draw [->] [thick] (3,1) -- (2.99,1.01);
	\draw (3,1) -- (1,3);
	\draw [->] [thick] (1,3) -- (1.01,2.99);
	\draw [->] (0,4) -- (1,3);
	\node at (-0.4,0.8) {$k_3$};
	\node at (-0.4,3.2) {$k_1$};
	\node at (4.4,0.8) {$k_4$};
	\node at (4.4,3.2) {$k_2$};
\end{tikzpicture}
\end{minipage}
\qquad 
V(k_1, k_2, k_3, k_4) 
=
- i \lambda \Bigl[ (\bk_1 \cdot \bk_2) (\bk_3 \cdot \bk_4) + (2\leftrightarrow3) + (2\leftrightarrow4) \Bigr].
$$
The one-loop correction to the coupling $\lambda$ comes from the candy diagram,
\begin{align} \label{eq:candyd}
& \quad \begin{minipage}{4cm}
\begin{tikzpicture}[scale=0.3]
	\draw [->] (0,0) -- (1,1);
	\draw [->] [thick] (1,1) -- (1.01,1.01);
	\draw (1,1) -- (2,2);
	\draw [->] (0,4) -- (1,3);
	\draw [->] [thick] (1,3) -- (1.01,2.99);
	\draw (1,3) -- (2,2);
	\draw [->] (4,2) circle (2cm);
	\draw [->] (8,4) -- (7,3);
	\draw [->] [thick] (7,3) -- (6.99,2.99);
	\draw (6,2) -- (7,3);
	\draw [->] (8,0) -- (7,1);
	\draw [->] [thick] (7,1) -- (6.99,1.01);
	\draw (7,1) -- (6,2);
	\node at (-0.5,4.8) {$k_1$};
	\node at (-0.5,-0.8) {$k_2$};
	\node at (8.5,4.8) {$k_3$};
	\node at (8.5,-0.8) {$k_4$};
	\draw [->,thick,domain=60:120] plot ({2.4*cos(\x)+4}, {2.4*sin(\x)+2});
	\node at (4,5.2) {$q = (\nu, \bq)$};
	\node at (4,-1.3) {$\phantom{q = (\nu, \bq)}$};
	\node at (4,6.3) {$\phantom{q}$};
\end{tikzpicture}
\end{minipage} 
\hspace{-8mm} =
\frac{1}{2} \int \frac{d\nu}{2\pi} \frac{d^2\bq}{(2\pi)^2} \, \Delta(q) \, V(k_1,k_2,q,-Q) \, \Delta(Q) \, V(k_3,k_4,-q,Q),
\end{align}
where
\be
    Q = (\nu + \omega_1 + \omega_2, \bq + \bk_1 + \bk_2),
\ee
Evaluating the candy diagram in \eqref{eq:candyd} and then summing over all channels give
\be 
    \frac{9\lambda^2}{4} \Big [ (\bk_1 \cdot \bk_2) (\bk_3 \cdot \bk_4) + (\bk_1 \cdot \bk_3) (\bk_2 \cdot \bk_4) + (\bk_1 \cdot \bk_4) (\bk_2 \cdot \bk_3) \Big ] \mathfrak{I}_2 + \text{finite},
    \label{eq:logdiv}
\ee
where
\be
    \mathfrak{I}_2 \equiv - i \int \frac{d\nu}{2\pi} \, \frac{d^2\bq}{(2\pi)^2} \, \frac{|\bq|^4}{(\nu^2 - |\bq|^4 - c^2 |\bq|^2)^2}.
\ee
Note that we expanded the integrand with respect to the smallness of the external frequency and momentum and only kept power-counting divergent contributions in \eqref{eq:logdiv}. Performing the Wick rotation $\nu \rightarrow i \nu$, we obtain
\be
    \mathfrak{I}_2 = \frac{1}{2\pi^2} \int d|\nu| \, d|\bq| \, \frac{|\bq|^5}{(\nu^2 + |\bq|^4 + c^2 |\bq|^2)^2}.
    \label{eq:mathfrakI}
\ee
When performing a frequency-momentum integral in $\mathfrak{I}_2$, it is prudent to change variables to the momentum raised to the power of $z = 2$ and perform the integral on the frequency-momentum${}^z$ plane. Let $\Omega$ and $\Lambda$ be some high frequency and momentum scales, respectively. Write the integral of interest as
\be \label{eq:xyI2}
    \mathfrak{I}_2 = \frac{\Omega}{4 \pi^2 \Lambda^2} \int dx \, dy \, \frac{x^2}{\bigl( \frac{\Omega^2}{\Lambda^4} y^2 + x^2 + \frac{c^2}{\Lambda^2} x \bigr)^2}\,,
\ee
where
\begin{align} \label{eq:xy}
    x &\equiv \frac{| \bq |^2}{\Lambda^2}, &%
    y &\equiv \frac{| \nu |}{\Omega},
\end{align}
and $R$ is the integration region bounded by the positive $x$ and $y$ axes and some curve in the positive quadrant of the integration plane. The integration region is not entirely arbitrary: it must be star-shaped with respect to the origin. In other words, given any point $(x,y) \in R$, the line segment connecting the origin to $(x,y)$ is contained in $R$. In this way, scaling $x$ and $y$ down equally (or, equivalently, scaling down frequency and momentum with dynamical critical exponent $z$), one never leaves the original integration region $R$. 

We can pass to polar coordinates $(r, \theta)$ with $0 \leq \theta \leq \frac{\pi}{2}$ by defining
\begin{align}
    x &= r \cos \theta, &%
    y &= r \sin \theta.
\end{align}
Since $R$ is star-shaped, the part of the boundary of $R$ which is not along the $x$ or $y$ axes can be parametrized in polar coordinates simply as
\begin{equation}
    r = f( \theta ),
\end{equation}
where $f ( \theta )$ is some single-valued positive function. We examine the behavior of the integral $\mathfrak{I}_2$ as we bring the cut-off surface radially closer to the origin by a small amount. Thus, we integrate out a thin shell of high energy modes contained in the region 
\be
	\CV = \Big\{ (r, \theta): \, b^z f (\theta) \leq r \leq f (\theta), \, 0 \leq \theta \leq \frac{\pi}{2} \Big\}, 
\ee
where $0< b < 1$ (the closer $b$ is to 1, the thinner the shell). Suppose that we are sufficiently close to the $z=2$ Gaussian fixed point that we can ignore the $c^2$ term. In other words, not only is $c^2 \ll \Lambda^2$, but we also integrate out a small shell of high-energy modes with $| \bq |^2 \gg c^2$. Then, \eqref{eq:xyI2} becomes
\begin{align}
    \mathfrak{I}_2 & = \frac{\Omega}{4\pi^2 \Lambda^2} \int_0^{\frac{\pi}{2}} d\theta \, \frac{\cos^2 \theta}{\lr \tfrac{\Omega^2}{\Lambda^4} \sin^2 \theta + \cos^2 \theta \rr^2} \int_{b^2 f(\theta)}^{f (\theta)} \frac{dr}{r} + \text{finite} \notag \\
        & = \frac{1}{8 \pi} \log \frac{1}{b} + \text{finite}\,.
\end{align}
Plugging the value of $\mathfrak{I}_2$ back into \eqref{eq:logdiv}, we obtain the one-loop quantum correction to $\lambda$ and therefore the beta function
\be
    \beta_\lambda \equiv \frac{d \lambda}{d \log (1/b)} = \frac{9 \lambda^2}{32 \pi} + \CO (\lambda^3).
\ee
Remarkably, this result is completely independent of the detailed form of $f (\theta)$. Moreover, no relation between the energy scale $\Omega$ and the momentum scale $\Lambda$ is required. It is clear that this argument also works for any theories with a definite dynamical critical exponent $z$ (at least for one-loop integrals).

\section{Bounds on Nonrelativistic Dispersion Relations}
\label{app:bounds}

In this appendix, we discuss the general method of applying bounds on Lorentz violations in experiments, which may be useful for testing the phenomenological viability of our proposal of nonrelativistic short-distance completion of the Higgs in the future. Reviews of  Lorentz violation tests can be found in~\cite{stefano13,Stecker:2017gdy,Kostelecky:2008ts} and references therein. 

There are many ways in which Lorentz symmetry violation can lead to observable physical effects. For instance, Lorentz invariance violating (LIV) terms can change the kinematics of particle interactions and decays, lower or raise the energy threshold of these processes, modify particle dispersion relations and even induce neutrino oscillations (see \cite{Stecker:2017gdy} and references therein). The LIV terms can be present in different sectors of a theory --- gauge, fermionic, scalar and gravity sectors. Lorentz invariance is also intricately related to \emph{CPT} invariance. Indeed, it was proven that if an interacting theory violates the discrete \emph{CPT} symmetry, it necessarily violates Lorentz invariance \cite{Greenberg:2002uu}. However, Lorentz invariance violation can happen with or without violation of \emph{CPT} symmetry.

The Standard Model extension (SME) framework was developed to explore systematically Lorentz violation \cite{Colladay:1998fq,Kostelecky:2003fs,Kostelecky:2009zp,Kostelecky:2011gq,Kostelecky:2013rta}. The SME is an effective field theory extension of the Standard Model coupled to general relativity with all possible LIV terms. A tremendous amount of work has been generated to classify and catalogue the constraints on all of these LIV terms using many experimental tests (see \cite{Kostelecky:2008ts}, which is updated annually).  

In the following, we will focus on a subset of these tests. We consider mainly the tests that probe modifications of particle dispersion relations due to LIV terms. In the nonrelativistic short-distance completion of the Higgs we proposed, we introduce \emph{CPT} invariant LIV terms in the scalar sector, \emph{i.e.}, the Higgs sector. Since the Higgs interacts with the fermions and the gauge bosons, LIV effects are communicated to these sectors as well.

The particle dispersion relation is often described in the literature as a power series in the energy $E$ of the particle,  \emph{e.g.},
\bea 
E^2 \approx |\bqp|^2 c^2 \left[ 1 + \sum \limits_{k=1}^{\infty} s_k \left(\frac{E}{M_\text{LIV}}\right)^k \right] \label{eq: dispersion_photon},
\eea
from which the particle speed $v(E) = \p E / \p |\bp|$ can be derived.
Here, $M_\text{LIV}$ is the LIV scale. For $s_k > 0$ ($s_k < 0$), the particle is superluminal (subluminal). The LIV scales for different particles may differ. We will denote by $M_a$, $M_f$ and $M$ the LIV scales associated with gauge bosons, fermions and the Higgs, respectively.

Experimental constraints on Lorentz violation in the Higgs sector have been considered only recently using ultrahigh energy cosmic rays~\cite{Cohen:2014tma}. In that work, the LIV comes from a single dimension-4 operator. By requiring that the cosmic ray particles (assumed to be protons) do not lose too much energy (\emph{e.g.}, through vacuum Higgs radiation), a constraint can be obtained on this LIV parameter. The derivation of the constraint demands going through the spontaneous symmetry breaking of $SU(2)_L \times U(1)_Y$ which is beyond the scope of this paper and is left for future study. Therefore, we turn now to experimental probes testing Lorentz violation effects on the fermion and gauge boson dispersion relations.

There have been many tests of Lorentz violating effects on the photon propagation from distant astrophysical objects, \emph{e.g.}, gamma ray bursts (GRB) \cite{Ellis:2005wr, RodriquezMartinez:2006xc,Bolmont:2006kk,Shao:2009bv,Nemiroff:2011fk,Vasileiou:2013vra,Chang:2015qpa} and active galactic nuclei (AGN) \cite{Albert:2007qk,Aharonian:2008kz}. These look for a time-of-flight difference between photons of different energies. The idea of using gamma ray bursts to put constraints on Lorentz violation was first raised in~\cite{AmelinoCamelia:1997gz}. Using these data (\emph{e.g.}, GRBs, AGNs), lower bounds on the LIV scale $M_\text{LIV}$ can be obtained. Bounds on $M_\text{LIV}$ can then be directly translated into bounds on the parameters of a theory given the photon dispersion relation in the theory. Therefore, we express the dispersion relation of the photon as in~\eqref{eq: dispersion_photon}. Keeping only the first higher-order correction to the photon dispersion relation,
\bea
E^2 &=& c^2 |\bqp|^2 + \zeta_{2,a}^2 |\bqp|^4 + \mathcal{O} ( | \bqp |^6 ) \nonumber \\
&\approx& |\bqp|^2 c^2 \left[ 1 + \tilde{\zeta}_{2,a}^2 E^2 + \mathcal{O}(E^4)  \right],
\eea
where $ \tilde{\zeta}_{2,a}^2 = \zeta_{2,a}^2/c^4$. In our model, higher-order dispersion terms for the photon are generated only indirectly via interactions with other particles which themselves interact with the Higgs (\emph{e.g.}, electrons) and are therefore highly suppressed. Taking $\tilde{\zeta}_{2,a}^2 \sim 1/M_a^2$ we obtain the photon dispersion relation 
\bea
E^2 &\approx& |\bqp|^2 c^2 \left[ 1 +  \frac{E^2}{M_a^2} + \mathcal{O} \left(\frac{E^4}{M_a^4}\right) \right].
\eea

The works \cite{Shao:2009bv,Vasileiou:2013vra,Chang:2015qpa} use GRBs to put a lower bound on $M_a$ for a quadratic dependence of the photon dispersion relation. The most stringent constraints come from \cite{Shao:2009bv,Vasileiou:2013vra}. The constraints derived are of the same order but the statistics were deemed insufficient due to the lack of data in \cite{Shao:2009bv}.  
The recent study~\cite{Vasileiou:2013vra} of GRBs detected by the Fermi Large Area Telescope sets $95\%$ lower bound on $M_a$
for a superluminal photon with a quadratic correction of $\sim 10^{7}$ TeV (see Table IV of~\cite{Vasileiou:2013vra}). The exact value of the lower bound depends on the GRB data and  the analysis method considered. This lower bound on $M_a$ can be directly translated into an upper bound on $\tilde{\zeta}_{2,a}^2$ if we assume that higher orders $E^n/M_a^n$ for $n>2$ are negligible.
Note that to derive bounds from GRBs data, $\Lambda$CDM is assumed.
The same order-of-magnitude bound on $M_a$ was found using observations by the MAGIC telescope of photons from active galactic nuclei \cite{Albert:2007qk}. 
 
Fermions with a higher-order dispersion relation (with the correct sign), and sufficiently high energy, will emit Cerenkov radiation and rapidly lose energy \cite{Stecker:2017gdy}. On the other hand, experiments have observed cosmic ray electrons directly with energy up to 5 TeV \cite{Staszak:2015kza} and indirectly with energy $\sim 100$ TeV from x-ray synchrotron radiation from supernova remnants \cite{Jacobson:2001tu, Koyama:1995rr}. This suggests that electrons cannot emit vacuum Cerenkov radiation below 100 TeV. A preliminary analysis in our model shows that electrons with energy below a few hundreds of TeV do not produce Cerenkov radiation. This is well above the current bound set by direct observation and is borderline with respect to the indirect bound. Note, however, that our estimate is conservative and can easily be improved.  

Lorentz violation in the fermionic sector can also 
be probed, for example, by studying neutrino oscillations in long-baseline experiments and time-of-arrival delay of neutrinos emitted by astrophysical sources such as supernova explosions.
In the work~\cite{Ellis:2008fc} the authors establish limits on Lorentz violation for neutrino dispersion relations using neutrino data from supernova $1987$a, data from Kamioka II, Irvine-Michigan-Brookhaven and Baksan. 

\bibliographystyle{JHEP}
\bibliography{sqed}

\end{document}